\documentclass[final,5p,times]{elsarticle}
\usepackage{graphicx}
\usepackage{amssymb}
\usepackage[amssymb]{SIunits}
\usepackage{pdfpages}
\usepackage{subfigure}
\usepackage{float}
\usepackage{amsmath}
\usepackage{color}
\usepackage{ulem}
\newcommand{\silvio}{Si$\Lambda$ViO\ }
\newcommand{\silvioo}{Si$\Lambda$ViO}
\newcommand{\lam}{$\Lambda$\ }

\begin{document}
\begin{frontmatter}
\title{Si$\Lambda$ViO: A Trigger for $\Lambda$-Hyperons}
\address[mue]{Excellence Cluster Universe, Technische Universit\"at M\"unchen,Boltzmannstr. 2, D-85748, Germany}
\address[darm]{GSI Helmholtzzentrum f\"{u}r Schwerionenforschung GmbH, Darmstadt, Germany}
\address[cler]{Laboratoire de Physique Corpusculaire, IN2P3/CNRS, and Universit\'{e} Blaise Pascal, Clermont-Ferrand, France}
\address[zagr]{Ru{d\llap{\raise 1.22ex\hbox{\vrule height 0.09ex width 0.2em}}\rlap{\raise 1.22ex\hbox{\vrule height 0.09ex width 0.06em}}}er Bo\v{s}kovi\'{c} Institute, Zagreb, Croatia} 
\address[heid]{Physikalisches Institut der Universit\"{a}t Heidelberg, Heidelberg, Germany}
\address[vien]{Stefan-Meyer-Institut f\"{u}r subatomare Physik, \"{O}sterreichische Akademie der Wissenschaften, Wien, Austria}
\address[sp]{University of Split, Split, Croatia}
\address[buda]{Wigner RCP, RMKI, Budapest, Hungary}
\address[wars]{Institute of Experimental Physics, Faculty of Physics, University of Warsaw, Warsaw, Poland}
\address[mosc]{Institute for Theoretical and Experimental Physics, Moscow, Russia}
\address[seou]{Korea University, Seoul, Korea}
\address[dres]{Institut f\"{u}r Strahlenphysik, Helmholtz-Zentrum Dresden-Rossendorf, Dresden, Germany}
\address[harb]{Harbin Institute of Technology, Harbin, China}
\address[kurc]{Kurchatov Institute, Moscow, Russia}
\address[buch]{Institute for Nuclear Physics and Engineering, Bucharest, Romania}
\address[stra]{Institut Pluridisciplinaire Hubert Curien and Universit\'{e} de Strasbourg, Strasbourg, France}
\address[tsing]{Department of Physics, Tsinghua University, Beijing 100084, China}
\address[rik]{RIKEN, Nishina Center, Wako, Saitama 351-0198, Japan}
\address[tok]{Deparment of Physics, University of Tokyo, Bunkyo-ku,Tokyo 113-0033, Japan}
\address[lan]{Institute of Modern Physics, Chinese Academy of Sciences, Lanzhou, China}
\author[mue]{Robert M\"unzer\corref{cor1}}
\author[mue]{Martin Berger\corref{cor1}}
\author[mue]{Laura Fabbietti\corref{cor1}}
\author[darm]{R.~Averbeck}
\author[darm]{A.~Andronic}
\author[cler]{V.~Barret}
\author[zagr]{Z.~Basrak}
\author[cler]{N.~Bastid}
\author[heid]{M.L.~Benabderrahmane}
\author[vien]{P.~Buehler}
\author[vien]{M.~Cargnelli}
\author[zagr]{R.~\v{C}aplar}
\author[sp]{I.~Carevic}
\author[wars]{V.~Charviakova}
\author[cler]{P.~Crochet}
\author[heid]{I.~Deppner}
\author[cler]{P.~Dupieux}
\author[sp]{M.~D\v{z}elalija}
\author[buda]{Z.~Fodor}
\author[mue,wars]{P.~Gasik}
\author[zagr]{I.~Ga\v{s}pari\'c}
\author[mosc]{Y.~Grishkin}
\author[darm]{O.N.~Hartmann}
\author[heid]{N.~Herrmann}
\author[darm]{K.D.~Hildenbrand}
\author[seou]{B.~Hong}
\author[darm,seou]{T.I.~Kang}
\author[buda]{J.~Kecskemeti}
\author[darm]{Y.J.~Kim}
\author[wars]{M.~Kirejczyk}
\author[darm,zagr]{M.~Ki\v{s}}
\author[mue,vien]{P.~Kienle\corref{cor2}}
\author[darm]{P.~Koczon}
\author[dres]{R.~Kotte}
\author[mosc]{A.~Lebedev}
\author[darm]{Y.~Leifels}
\author[heid,harb]{J.L.~Liu}
\author[cler]{X.~Lopez}
\author[kurc]{V.~Manko}
\author[vien]{J.~Marton}
\author[wars]{T.~Matulewicz}
\author[buch]{M.~Petrovici}
\author[wars]{K.~Piasecki}
\author[stra]{F.~Rami}
\author[heid]{A.~Reischl}
\author[darm]{W.~Reisdorf}
\author[seou]{M.S.~Ryu}
\author[vien]{P.~Schmidt}
\author[darm]{A.~Sch\"{u}ttauf}
\author[buda]{Z.~Seres}
\author[wars]{B.~Sikora}
\author[seou]{K.S.~Sim}
\author[buch]{V.~Simion}
\author[wars]{K.~Siwek-Wilczy\'{n}ska}
\author[mosc]{V.~Smolyankin}
\author[vien]{K.~Suzuki}
\author[wars]{Z.~Tyminski}
\author[stra]{P.~Wagner}
\author[vien]{E.~Widmann}
\author[heid,wars]{K.~Wi\'{s}niewski}
\author[tsing]{Z.G.~Xiao}
\author[rik,tok]{T.Yamazaki}
\author[kurc]{I.~Yushmanov}
\author[lan]{X.Y.~Zhang}
\author[mosc]{A.~Zhilin}
\author[heid]{V.~Zinyuk}
\author[vien]{J.~Zmeskal}
\cortext[cor1]{Corresponding author}
\cortext[cor2]{deceased}

\begin{abstract}
    As online trigger for events containing $\Lambda$ hyperons in p+p collisions at $\mathrm{3.1\,GeV}$ a silicon-based device has been designed and built. This system has been integrated close to the target region within the FOPI spectrometer at GSI and was also employed as a tracking device to improve the vertex reconstruction of secondary decays. The design of the detector components, read-out, the trigger capability as well as the tracking performance are presented. An enrichment factor of about $14$ was achieved for events containing a $\Lambda$-hyperon candidate.
\end{abstract}


\end{frontmatter}

\section{Introduction}
A modular silicon-based system has been designed to trigger on the particular signature of $\mathrm{\Lambda}$ hyperons decaying into $\mathrm{p-\pi^-}$ pairs. Similar concepts have been developed for the DISTO \cite{DISTO} and COSY-TOF \cite{COSYTOF} experiments exploiting scintillating fibers arranged in several layers and placed close to the target regions. There, the trigger was based on different hit multiplicities in the various layers of the scintillating fibre detectors. We have  utilized single and double-sided, partially high-segmented silicon detectors as a trigger device. The setup was designed to reduce dead areas and improve the hit position resolution. This way we have combined a rather fast online computation of the hit multiplicity pattern, used to build the second level trigger, with an improved position resolution of the order of 1 mm.
This device was named \silvio (\underbar{Si}licon \underbar{$\mathrm{\Lambda}$} \underbar{V}ertexing and \underbar{I}dentification \underbar{O}nline) designed to fit within the central tracking chamber of the FOPI \cite{FOPI1} spectrometer at GSI. Indeed, the final goal of this development is to search in the reaction $\mathrm{p+p}$ $\mathrm{\rightarrow p+\Lambda+K^+}$ for the formation of so-called $\mathrm{ppK^-}$ bound state together with a $\mathrm{K^+}$ leading to the final state \cite{PROPOSAL}. Theoretical calculations \cite{Yam1,Yam2} suggest that the formation of such state is favoured at the beam energies available at GSI, i.e. max 4.5 AGeV. Moreover, the large geometrical acceptance of the FOPI spectrometer \cite{FOPI1} fulfills requirements for this exclusive measurement.
Apart from a significantly enhancing the amount of events with Lambda candidates \silvio allows for vertex reconstruction of tracks emitted below the polar angle of $27^{\circ}$ which is lacking in the FOPI setup \cite{LOPEZ}.
 This present paper is organized as follows: section one shortly describes the FOPI apparatus, section two summarizes the \silvio detector concept, its schematics and read-out. In section three the calibration of the signals is discussed, followed by the trigger performance in section four. In sections five and six the tracking algorithm, efficiency and purity are shown together with the reconstruction of the $\mathrm{\Lambda}$ hyperons and the selection capability of the trigger, based on both full-scale simulations and experimental data. We conclude with a summary in chapter seven.
\subsection{The FOPI spectrometer}
FOPI \cite{FOPI1,FOPI2,FOPI3} is a fixed target spectrometer located at GSI (Darmstadt, Germany).
It consists of an inner drift chamber which is surrounded by a scintillator and a RPC time-of-flight barrel \cite{RPC1,RPC2}.
A second drift chamber and a scintillator time-of-flight wall are placed in the forward direction.
The drift chamber and the time of flight barrels are surrounded by a superconducting \unit{0.6}{\tesla} solenoid magnet.
FOPI has nearly a 4$\mathrm{\pi}$ geometrical acceptance, a momentum resolution between 7\%-10\%, a time-of-flight resolution between 80 ps (RPC) and 200 ps (plastics scintiallators) in heavy ion reactions depending on the detector type and a vertex resolution of \unit{5}{\milli\meter} in the XY-plane for tracks measured with the Central Drift Chamber (CDC). 
FOPI was designed as a heavy ion spectrometer and is therefore not best suited to trigger on low multiplicity elementary reactions. The \silvio system enables to trigger on
reactions from proton-proton collisions where a $\mathrm{\Lambda}$ is produced in the final state.\\
Moreover, the poor spatial resolution of the forward tracking detector HELITRON does not allow for vertex reconstruction and identification of secondary vertices by itself. \silvio can improve the tracking capabilities especially in the forward direction by providing an additional hit-point for charged particle tracks close to the target region. 
\subsection{Beam Line Detectors}
\begin{figure}[ht]		
\centering	
\includegraphics[width=.9\columnwidth]{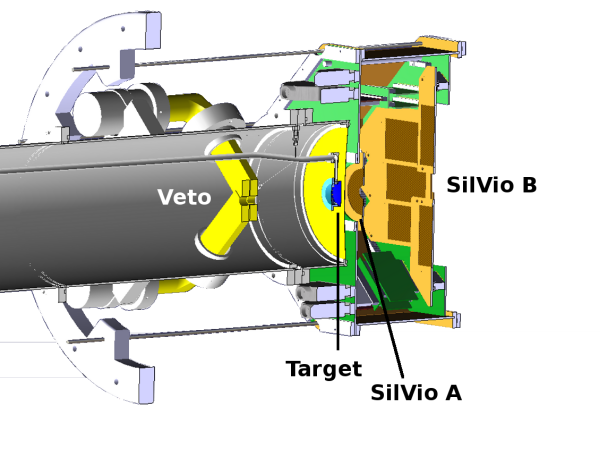}
\caption{Crossection of the target area with veto-detector, target cell and \silvio.}
\label{pic:targetregion}
\end{figure}

In addition to the \silvio system, two new detectors for the beam characterization have been developed:\\
The start detector - placed around \unit{2}{\meter} upstream of the target - was made out of five \unit{10\times2}{\centi\square\meter} and \unit{10}{\milli\meter} thick scintillator plates. 
To cope with beam intensities up to \unit{10^{7}}{protons/s}, focused on a beam spot of about $6\,\mathrm{mm}$ diameter, the readout was done on both sides with booster photo multipliers\footnote{Hamamatsu H6524-01MOD with modified booster lines}, which works to rates up to several \unit{}{\mega\hertz}. Since the magnetic field inside FOPI at the position of the start detector is still in the order of \unit{60}{\milli\tesla} the photo multipliers were shielded with iron tubes. The intrinsic time resolution of this detector was about \unit{130}{\pico\second} at a hit rate of \unit{2~}{\megad\hertz} \cite{beam}, which leads to a time-of-flight resultion of 130 ps for the RPC and 215 ps in the plastic barrels.\\
Beam particles undergoing upstream interactions or stemming from a beam halo were suppressed using a veto-detector.
The veto-detector consists of two \unit{5}{\milli\meter} thick scintillators which were read out on one side only by fine-mesh PMTs \footnote{Hamamatsu H6152-01B with modified booster lines}. 
The two scintillators are rotated by \unit{90}{\degree} with respect to each other and were split into two segments each, to facilitate the installation. Both detectors have an inner hole of \unit{1}{\centi\meter} diameter to enable the passage of the beam and were positioned around \unit{10}{\centi\meter} before the target edge.
The efficiency of this detector was evaluated to be above 98\% \cite{beam}. 
Figure~\ref{pic:targetregion} shows the veto detector together with the target inside the evacuated beam tube and the \silvio setup.
\subsection{Target}
Liquid hydrogen contained in a cell of kapton was used in the experiment. The target cell was made of a \unit{2}{\centi\meter} tube with a diameter of \unit{3}{\centi\meter} mounted in the evacuated beam pipe.
Hydrogen was cooled down by a cryogenic cooler outside the setup. \\
The resulting interaction probability of protons with a beam energy of $\mathrm{3.1\,GeV}$ with the target structure was 0.004\% for the capton foil and 0.372\% for the liquid hydrogen.
\section{The $\mathrm{\Lambda}$-Trigger}
\silvio (\underbar{Si}licon-\underbar{$\mathrm{\Lambda}$}-\underbar{V}ertexing-and-\underbar{I}dentification-\underbar{O}nline)
is a detector system designed to generate a trigger signal based on the decay pattern of a \lam hyperon and to improve the tracking capabilities of FOPI for particles emitted in the forward direction.
\silvio is composed of two layers of silicon detectors, hereafter referred to as \silvioo-A and \silvioo-B.\\
The basic idea of the \lam trigger is that the multiplicity in two consecutive layers of detectors is different when a $\mathrm{\Lambda}$ decays between them.
In this case, since the $\mathrm{\Lambda}$ is a neutral particle, no signal is generated in the first layer (\silvioo-A) but two charged particles are detected by the second layer (\silvioo-B).
The precise acceptance is determined by the combined analysis of simulated and experimental data.
\begin{figure}[ht]
\centering
\includegraphics[width=0.90\columnwidth]{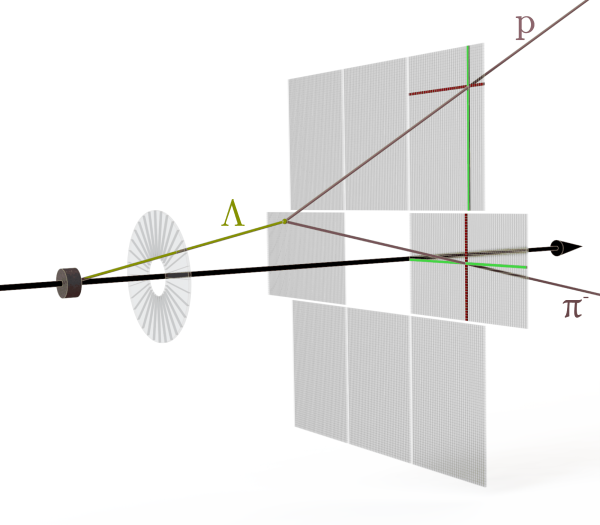}
\caption{Schematic view of the working principle with an decaying \lam . The detectors are drawn in grey while sectors/strips with a generated signal are drawn in green or red.}
\label{pic:silvioB} 
\end{figure}\subsection{Detector Setup}
\subsubsection{Silvio A}
\silvioo-A is a single sided annular \unit{1}{\milli\meter} thick silicon detector divided into 32 sectors with an inner and outer radius of 7 and \unit{23.5}{\milli\meter} respectively. 
The detector itself is mounted on a small PCB which can be attached to the \silvioo-B main board.
This main board, not only holds the detector, but also includes all the necessary feedthroughs and connectors for the readout. 
Positioned \unit{3}{\centi\meter} downstream from the target the detector covered a polar angle ranging from \unit{13}{\degree} to \unit{38}{\degree}. The thickness of the device was selected to make sure that the charge deposited by the minimal ionising particle (MIP) is sufficient to separate this signal from the electronic noise and low energy $\mathrm{\delta}$-electrons.
\subsubsection{Silvio B}
\begin{figure}[h]
\centering
\includegraphics[width=0.9\columnwidth]{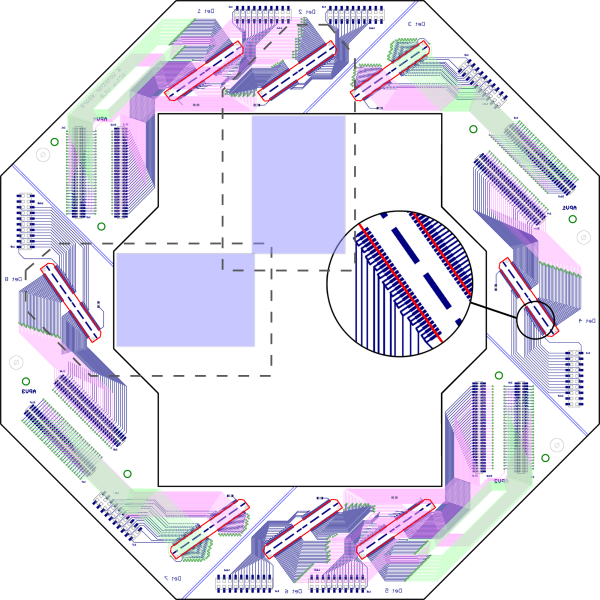}
\caption{\silvio B detector PCB. The inlet shows how the strips of one detector were combined for the readout.
The positions of two detectors are marked as a grey dashed line with their active areas in light blue.
All the connectors for the detectors are marked in red.}
\label{pic:sBboard}
\end{figure}

\silvioo-B is a patchwork of 8 double sided silicon strip detectors with a pitch of \unit{1}{\milli\meter} and 
a thickness of \unit{1}{\milli\meter}. 
Each detector has a size of \unit{40x60}{\milli\square\meter} with 60 strips on the p-side, 16 strips on the n-side. The individual detector PCBs - optimized for a minimum of passive material - are mounted on a common main board.
This configuration guarantees a good position resolution in the polar angle direction, while the detector side, corresponding to the azimuthal angle, has a rougher binning which suits the trigger realization. 
Figure~\ref{pic:sBboard} shows the layout of the \silvioo-B main board.
The positions of two detectors are exemplary shown in gray dashed line with the active area in light blue and their connectors marked in red.
The inlet shows how the previously mentioned grouping of n-side readout strips is realized.
All detectors were produced by Canberra \cite{canb}.
\subsubsection{Detector Box}
\label{sub:detbox}
\begin{figure}[H]
\centering
\includegraphics[width=\columnwidth]{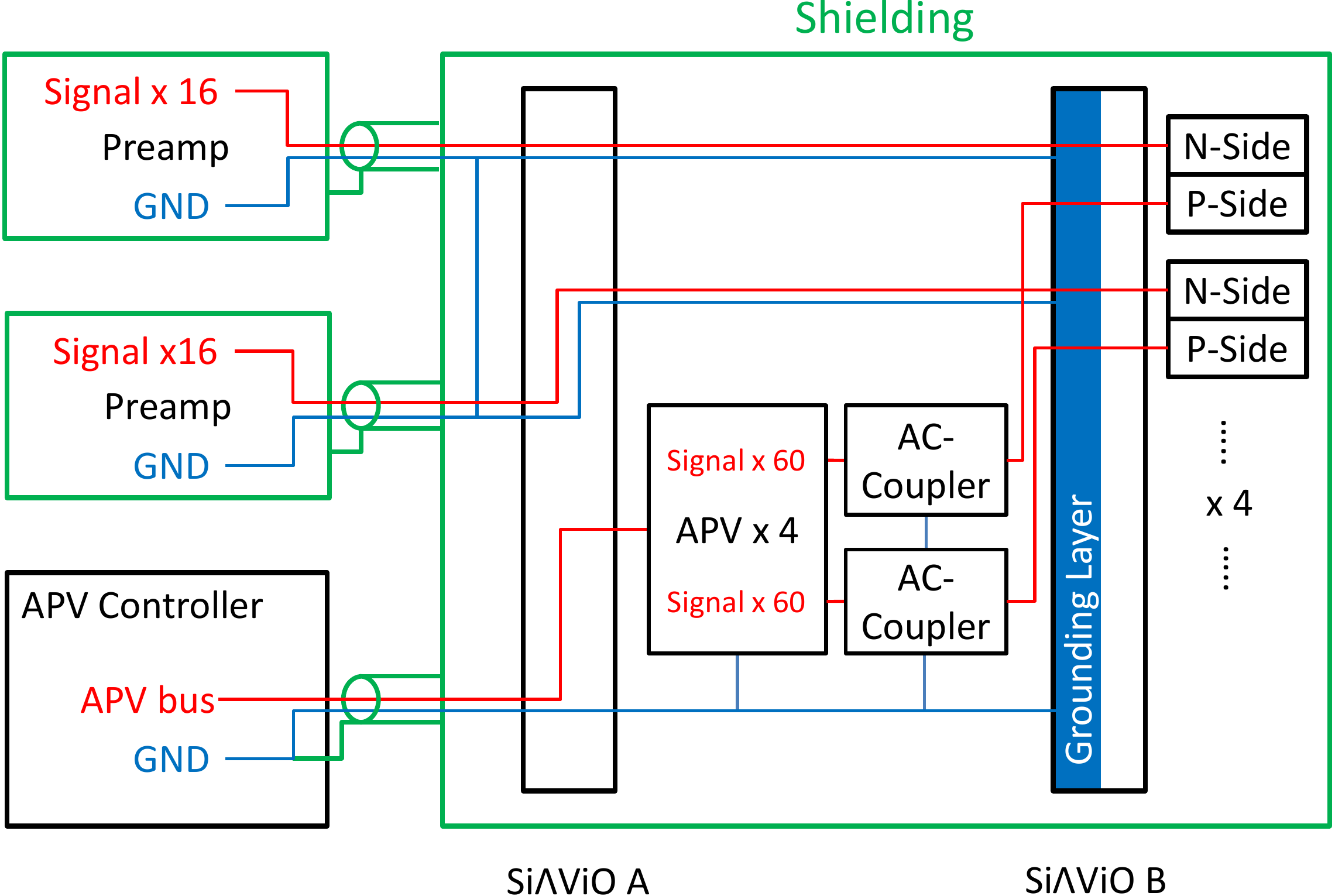}
\caption{Schematic view of the grounding and shielding concept.}
\label{pic:gnd_shield}
\end{figure}

The two main boards of the trigger detector are mounted in a fixed distance between the two detector plains.
The p-side of the \silvioo-B is read out by APV combined with an AC-Coupler for the charge attenuation (see section~\ref{subchap-Readout}). These are mounted on boards positioned between the two layers, which allows for a quite short length of the readout lines from the p-side strips of the detector to the readout chip.\\
\begin{figure}[H]
\centering
\includegraphics[width=\columnwidth]{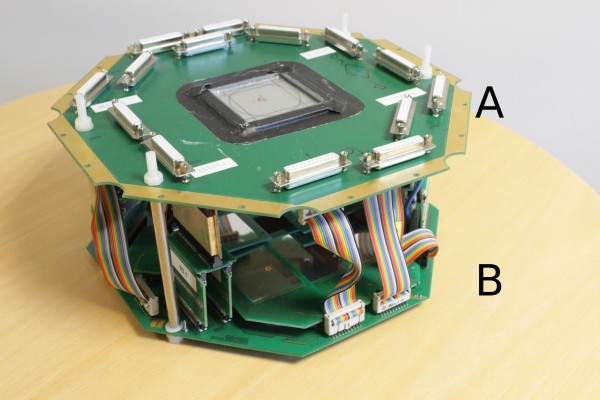}
\caption{The \silvio detector without the shielding box. On the bottom side one can see the \silvioo-B layer with the eight rectangular detectors. On the upper part the front side of the \silvioo-A PCB can be seen, which holds the connectors and serves as a shielding.}
\label{pic:silviopic}
\end{figure}
The signals of the n-side strips are lead by flat cables to the \silvioo-A PCB, where they are fed to sub-d connectors. 
All signals are collected on the top side of the \silvioo-A main board.
This setup can be seen in figure~\ref{pic:silviopic}.
In total there are 16x8 channels from n-side and 60x8 channels from the p-side that are transferred from the \silvioo-B board to the \silvioo-A board.
The p-side signals, however, are first sent to the APV25 board where they are processed and further transferred in a multiplexed way.
This drastically reduces the amount of signal cables.\\
Both main boards have cutouts in correspondence of the active detector areas to reduce the radiation length.
To protect the detector from light and for shielding against electrical interferences the setup is put into a hexagonal 
shaped box made out of epoxy laminated paper, coated with a thin copper layer. The upper side is closed by the \silvioo-A PCB. 
The bottom side is closed by an \unit{0.1}{\milli\meter} thick aluminum foil, having a complete shielding as shown in figure~\ref{pic:gnd_shield}.
Shielded cables were used for all signals to avoid noise pickup. The shielding allows a trigger threshold as low as 220 keV (see section~\ref{chapt:sign:raw})
\subsubsection{Readout Electronics}
\label{subchap-Readout}
\begin{figure}[ht]
\centering
\includegraphics[width=\columnwidth]{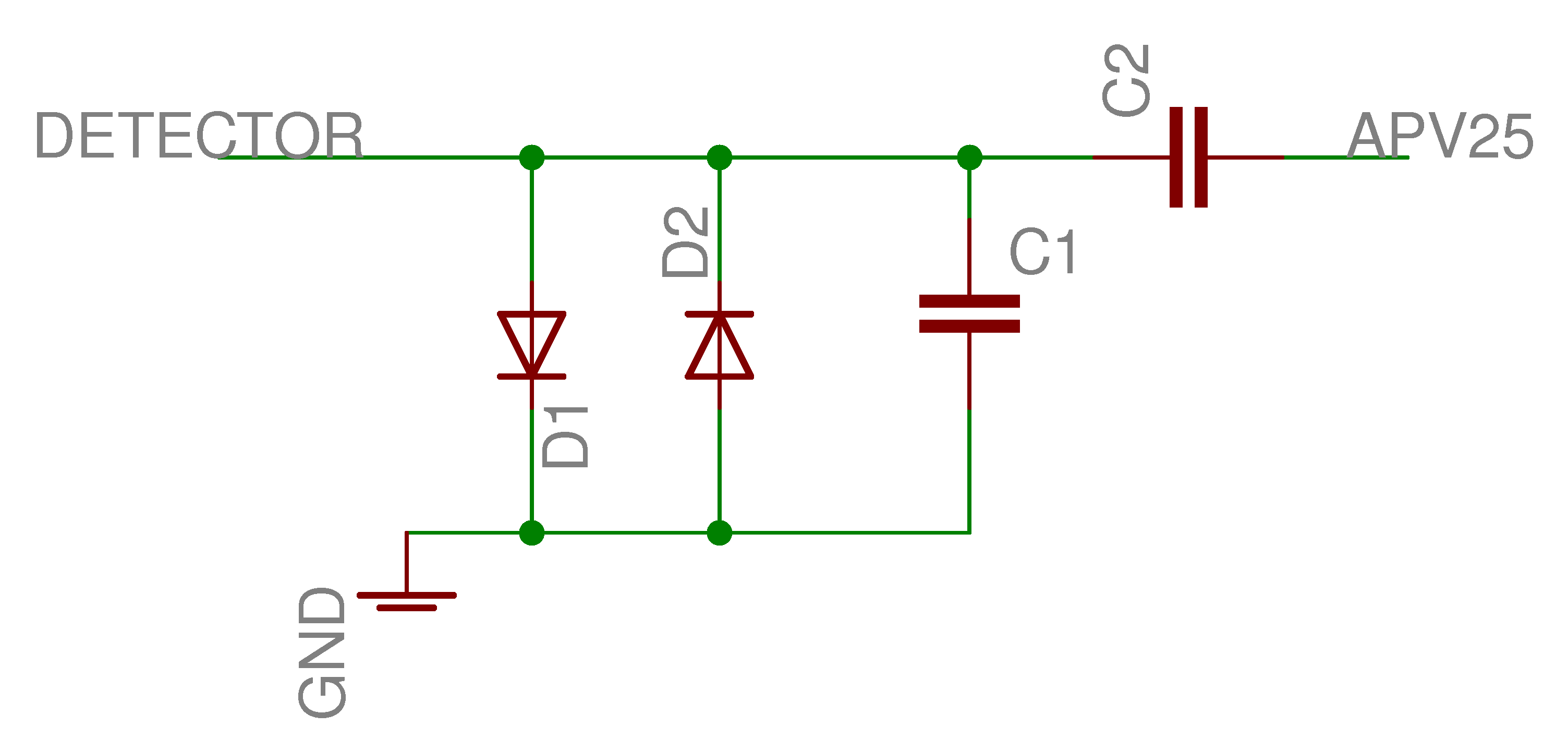}
\caption{The schematics of the AC-Coupler with charge divider (C1 and C2) and protection circuit (D1 and D2).}
\label{pic:ac-coupler}
\end{figure}

For the readout two different systems are used. Both systems are able to process positive and negative signals. The signals of the segmented side \silvioo-A and the n-doped side of \silvioo-B are processed with analog preamplifier\footnote{Mesytec MPR16\cite{mesytec}} with a range of \unit{5}{\mega\electronvolt}. The amplified signals are fed into STM16+ NIM Moduls from Mesytec. This shaping module has a fast timing filter amplifier for generating the trigger signal based on a user-defined and variable individual threshold for each channel and a slow shaper with \unit{2}{\micro\second} for the amplitude signal.
These modules generate a trigger signal if the hit multiplicity is within an interval defined by the user.
Each shaper has 16 inputs and it is possible to daisy chain more devices together to generate a signal on the base of a global multiplicity.
Furthermore this signal - used to determine the global multiplicity - is read out by a QDC \footnote{CAEN v985}. A QDC had to be used because this signal is current based. The signal of the shapers has to be attenuated to fit to the QDC input range.\\
The shaped signals are then further processed by 12-bit peak sensing ADCs \footnote{CAEN v786}. 
The p-doped side of the \silvioo-B detectors is read out by front-end electronics, based on the development for the HADES-RICH upgrade, equipped with the APV25 chips \cite{APV3,APV4}. 
The APV25 has 128 input channels, each with a preamplifier and a shaper with \unit{50}{\nano\second} shaping time (variable between 30 and \unit{400}{\nano\second}). 
These signals are sampled at at frequency of \unit{40}{\mega\hertz} and stored in a 192 cell ring buffer. The total length of the ring buffer of \unit{4.8}{\micro\second} is sufficient for the trigger latency.\\
Since each detector has 60 strips on the p-side, one APV25 with 128 channels was used for the readout of 2 detectors.
The 8 remaining channels are distributed evenly among the 120 signal channels to allow for an online FPGA based common mode noise correction.
\begin{figure}[ht]
\centering
\includegraphics[width=.9\columnwidth]{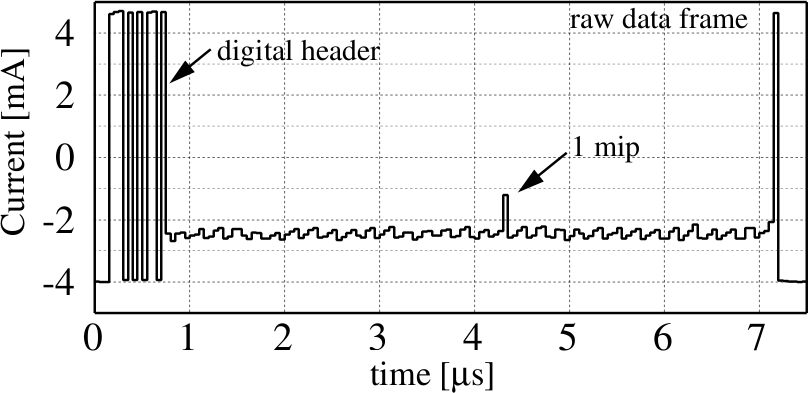}
\caption{A complete output data frame of the APV25 chip following a trigger with a the signal of one MIP in one channel\cite{APV4}. The digital header is followed by the multiplexed analog data and ends with an synchronisation pulse.}
\label{pic:apv_logic}
\end{figure}
After each trigger the amplitude of all 128 channels are multiplexed to the output stage. Figure~\ref{pic:apv_logic} shows one output data frame of the APV25. 
The output data are digitized and sent to a SAM3 \cite{gsiweb} VME module via a dedicated backend board \cite{bridgeboard}.  
In order to choose the right cell of the buffer to be read out one can set a fixed latency time. The procedure to find the correct latency in the experiment is described in section~\ref{chapt:soft:latency}. 
In order to be able to measure also particles with a much larger energy depositions than MIPs a capacitive charge divider was used.
Figure~\ref{pic:ac-coupler} shows the schematic of this charge divider (C1=\unit{47}{\pico\farad} and C2=\unit{20}{\pico\farad}) as it was implemented for every channel. 
Furthermore a protection circuit (D1 and D2) was embedded.
\begin{figure}[ht]
\centering
\includegraphics[width=\columnwidth]{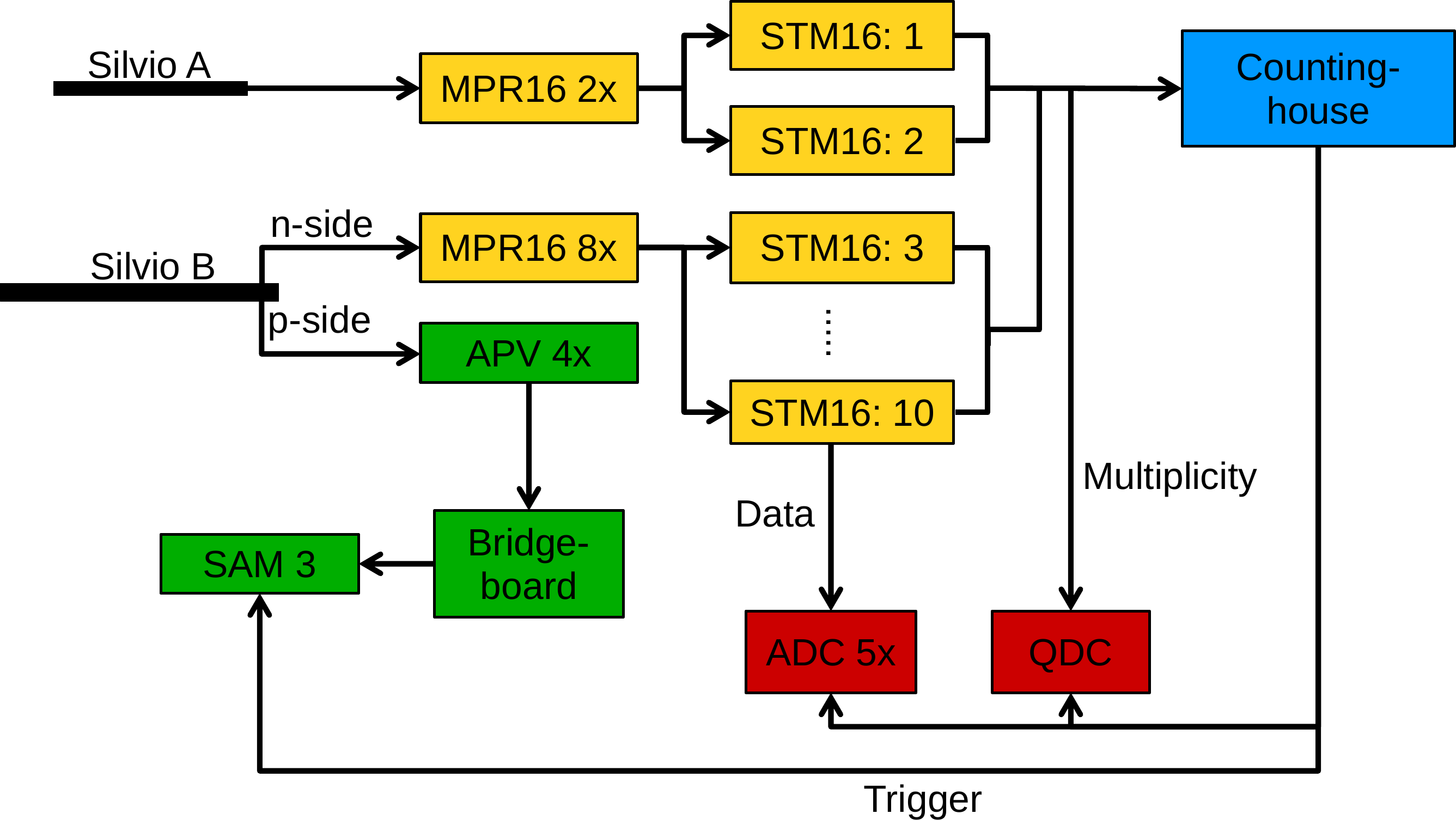}
\caption{The readout structure for \silvio.}
\label{pic:apvreadoutchain}
\end{figure}
Figure~\ref{pic:apvreadoutchain} shows the full readout scheme.
The signals from \silvioo-A and from the n-side of \silvioo-B are processed with the analog Mesytec electronics to generate the trigger signal which was sent to the main trigger crate.
Here this signal was logically combined with all the other trigger signals from the FOPI detectors.
This whole trigger generation procedure together with cable delays took \unit{1}{\micro\second}.
\section{Detector Signals}
\subsection{Raw Spectra}
\label{chapt:sign:raw}
\begin{figure}[ht]
\begin{picture}(128,346)
  \put(0,0){\includegraphics[width=0.98\columnwidth]{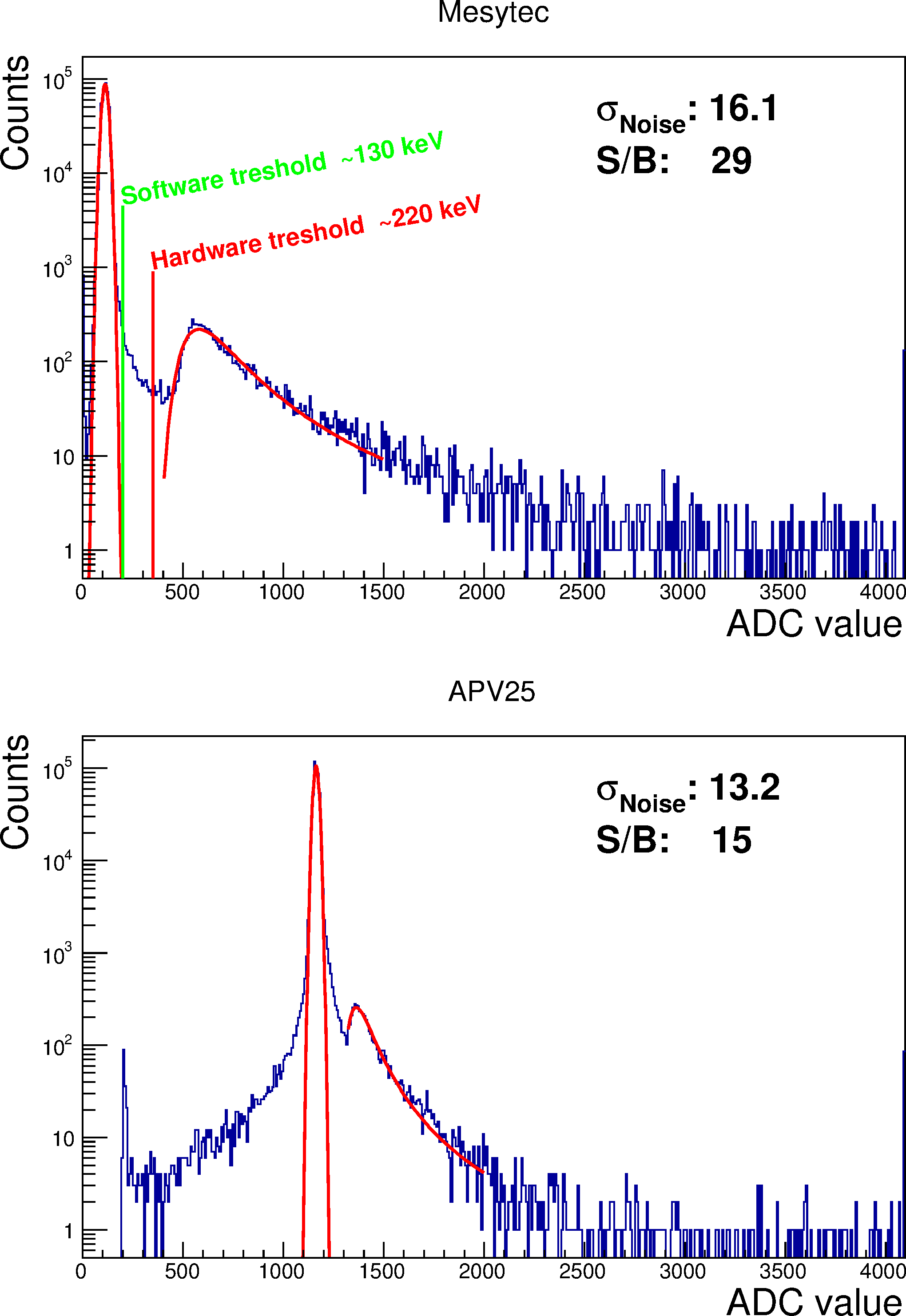}}
  \put(160,218){(a)}
  \put(160,60){(b)}
\end{picture}
  \caption{Raw Spectra of one channel for the Mesytec readout (a) and for the APV25 (b). The left peaks in each spectrum, which are fitted by an Gaussian function are the noise distribution, the left peaks are the signals of MIPs. The positions of the hardware/software threshold are marked with the red/green line (see text)}
  \label{pic:rawsig}
\end{figure}

The raw signal distributions for the two different readout electronics are shown in Figure~\ref{pic:rawsig}.
 The amplitude distribution for one channel of the analog Mesytec (a) and of the APV25 (b) are shown in Figure~\ref{pic:rawsig}.
One can clearly see the MIP peak (right peak) well separated from the noise peak (left peak) for both cases.
A signal to noise ratio (S/B) of 29 and of 15 was obtained for the Mesytec and for the APV25 read-out respectively. The events below the noise peak in the APV25 spectrum are caused by the drop of the baseline (see section~\ref{sec:basecorr}).\\
The hardware threshold was set for each channel separately directly above the noise level.
In the offline analysis a threshold corresponding to 3.5 times the noise width was chosen.\\
The hardware and the software threshold are shown in figure~\ref{pic:rawsig} (green line and red line respectively). The energy values were deduced from the calibration with the known energy loss of MIP in 1mm silicon. The energy value of the hardware threshold is deduced from the threshold, which is correlated to the energy.\\
The values below the noise peak in the APV25 raw spectra (figure~\ref{pic:rawsig}) are stemming mainly from rare events where a baseline drop occurred (see section~\ref{sec:basecorr}).
\subsubsection{Latency scan}
\label{chapt:soft:latency}
\begin{figure}[ht]
\centering
\includegraphics[width=.9\columnwidth]{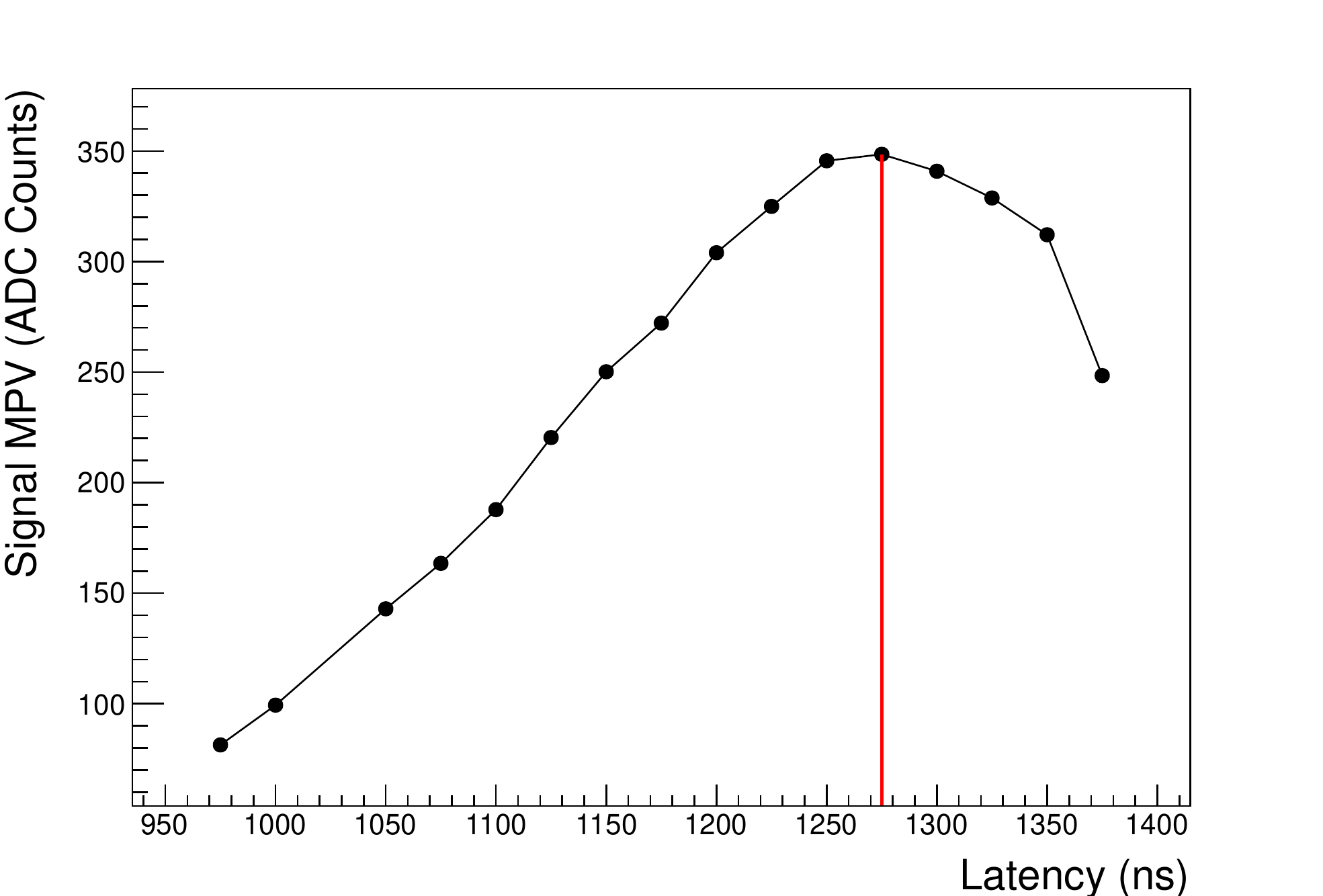}
\caption{Most probable value of signal amplitude as a function of latency in ns. The maximal amplitude is at a latency value of \unit{1275}{\nano\second}}
\label{pic:soft:latency2}
\end{figure}

To optimize the signal to background ratio of the APV25 the latency between the internal shaping in the chip to the readout trigger had to be determined and set in the chip, which can be done in units of the sampling clock ( \unit{25}{\nano\second} ).\\
In the setup only one sample is read, which increases the readout speed of the APV25 at the expense of slightly decreasing energy resolution.\\
The latency was determined by the following procedure:\\
 Around a starting value, derived from the cable length, delays inside the trigger logic and pulser measurement, different latency values were set, each time taking a file with a couple of minutes of beam data. 
For each setting the most probable value (MPV) of a Landau fit of the MIP signal, summed over all strips, was determined.
Figure~\ref{pic:soft:latency2} shows the MPV as a function of the set latency value in the APV25. A latency of 1275 ns ( 51 clock cycles) was found to have the highest amplitudes, which is well below the maximum size of the ring buffer.
\subsubsection{Baseline Correction}
\label{sec:basecorr}
\begin{figure}[ht]
\centering
\includegraphics[width=.8\columnwidth]{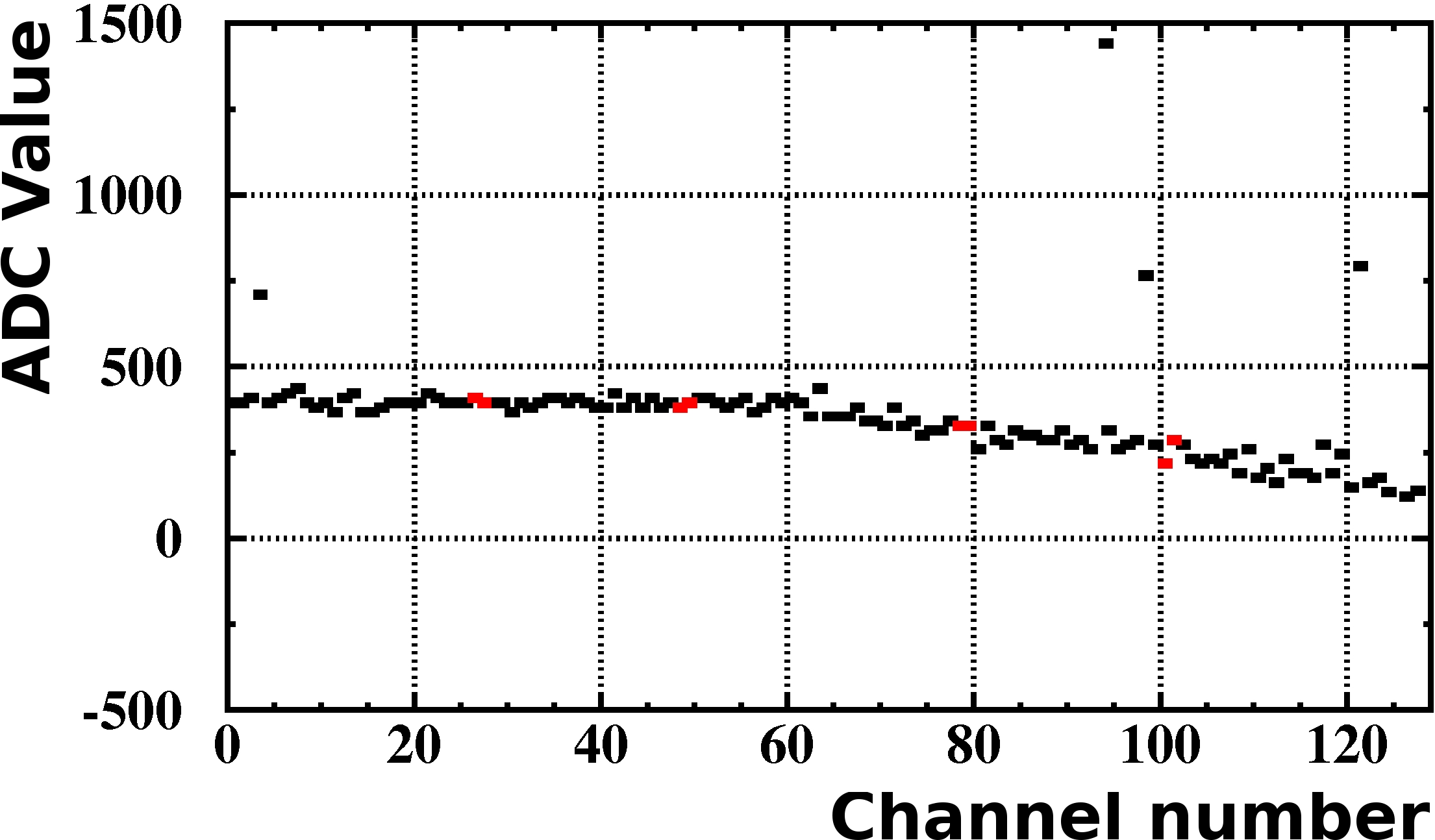}
\caption{ADC value plotted versus the channel number for one event and APV25 chip. A drop of the baseline is clearly visible on the right side. The red marked channels are unconnected and could be used for an online FPGA based correction of this effect.}
\label{pic:soft:bdrop}
\end{figure}

The powering scheme of the APV25 causes particular effects on the offset of the individual channels. In case the power consumption is high, due to either one channel with large amplitude or several hit channels, the baseline shifts to lower values with a special pattern.
Figure~\ref{pic:soft:bdrop} shows the ADC value plotted versus the channel number of one APV25 chip. For an event with 3 charges particles hitting the corresponding detector a drop of the baseline is clearly visible.
\\
Since in this experiment no zero suppression was applied, the correction for the baseline shift effect can be performed off line. 
The correction is done by fitting a polynomial function of 5th order to the amplitude vs channel number histogram.
Then the data points are shifted corresponding to the fit results in order to restore a common baseline.
\subsubsection{Multiplicity -Signals}
\begin{figure}[ht]
\centering
\includegraphics[width=.8\columnwidth]{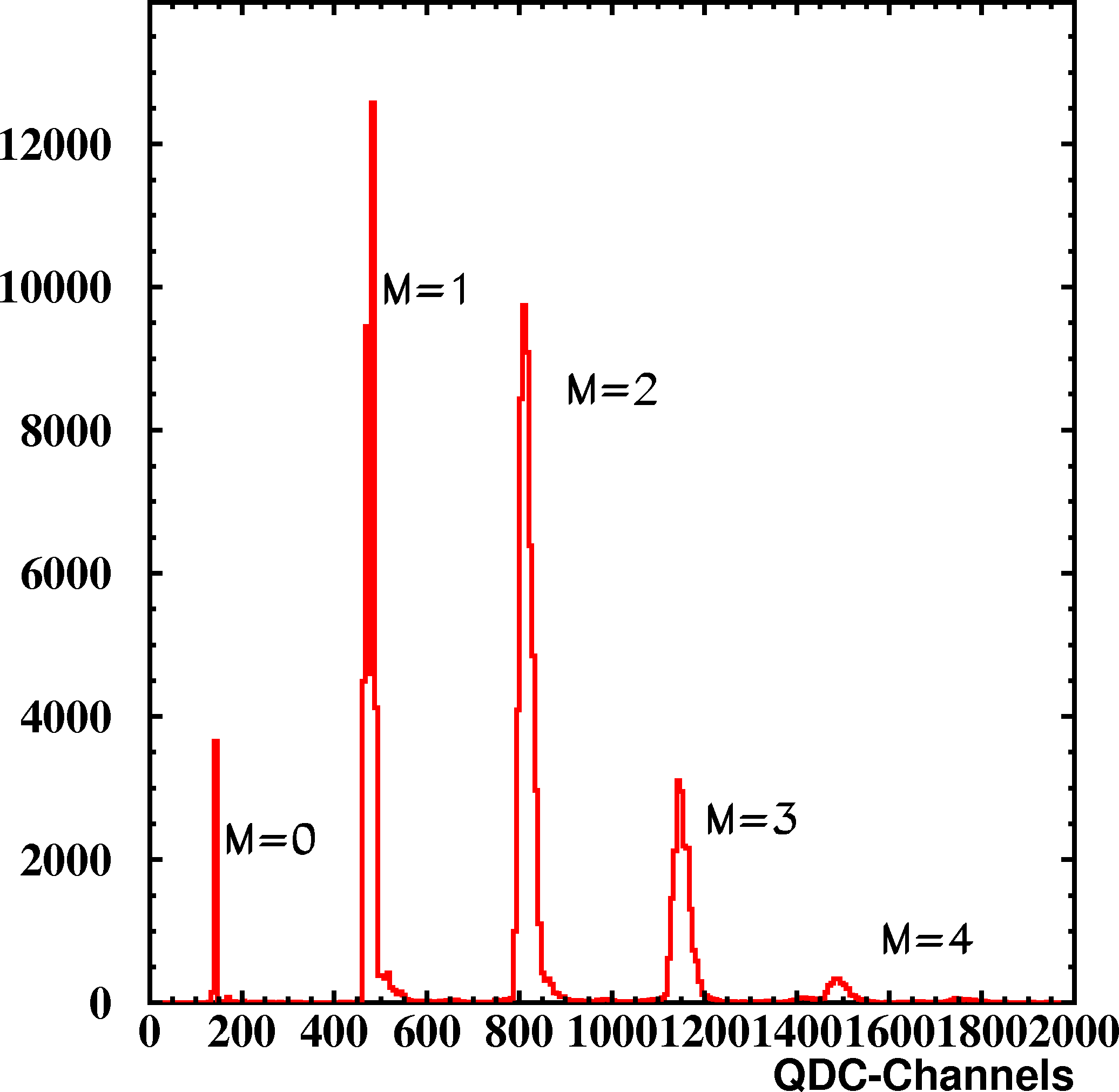}
\caption{Multiplicity Signal for \silvio B layer, with the assigned multiplicity values M from 0 to 4 hits.}
\label{pic:soft:curroutmult}
\end{figure}

The multiplicity signal from the STM16+ shapers is used to record the trigger hit multiplicity of the detector layers.
Due to electronic noise, these signals are widened. A clear correlation between signal and hit multiplicity can only be made, if well separated peaks are visible, which can be assigned to different multiplicities. Figure~\ref{pic:soft:curroutmult} shows the spectrum of the multiplicity signal of the \silvioo-B layer. The different peaks, which are labeled with the corresponding multiplicities, are quite well separated.\\
The corresponding multiplicity is selected by cutting around those peaks. The impurities of those cuts are below 1\%.
\section{Test of Trigger Electronics}
\label{TrigPer}
\subsection{Test of Trigger Electronics}
\subsubsection{Purity and Efficiency of Electronics}
In a dedicated experiment at the FOPI beam line the purity and the efficiency of the trigger electronics was tested. The trigger electronics purity is defined as the fraction of events, in which the selected hardware multiplicity equals the number of reconstructed particles hits. This number is determined in an offline analysis by counting the number of detector strips with a signal higher then the software threshold.\\
In this experiment two different detectors were used to measure the purity for multiplicities equal 1 and 2 at the same time \cite{rmue}.\\
The resulting purity values are $94.8 \pm 0.5 \%$ and of  $98.3 \pm 2.2\%$ for the multiplicity conditions one and two respectively.
The trigger electronics efficiency is defined by the ratio of single particles passing the detector resulting in a positive trigger response divided by the total number of hits.\\
To reconstruct the particles passing through the detector, two additional tracking detectors were used. \cite{rmue}\\
After disentangling the detector efficiency, the trigger electronics efficiency was determined to be $97.1 \pm 0.4$ \% \cite{rmue}.
\subsubsection{Charge Sharing and Clustering}
The measured particle multiplicities is disturbed by charge sharing between neighboring strips, induced by cross talk or particle, which do not cross the detector perpendicular to its surface.\\
In a separate test at the the MLL Tandem accelerator in Garching (TUM) \cite{MLL} a low intensity proton beam was directed onto the detector at different impact angles from 0\degree to 30\degree. For each angle the ratio of events with two or three neighboring strips, in which a signal was induced, was measured.\cite{rmue}\\
The total influence of the amplitude sharing was determined by folding the charge sharing curve with the expected angular distribution in the experiment, which is $8.20\pm 0.20 \%$.
Three strip charge sharing could be neglected in this angular region \cite{rmue}. \\
In order to combine the shared charge of two or more neighboring strips a cluster algorithm was employed. This algorithm simply combines neighboring 
strips on one detector side which were above threshold and calculates the cluster position by the amplitude-weighted mean of the strip positions. The mean number of strips per cluster for the p- and n-side are $2.637$ and $1.247$, respectively.\\
After the clustering, the two detector sides are combined. Every cluster on the p-side is combined with every clusters on the n-side to hitpoint candidates. To determine which combination is the correct one the track information of the other FOPI detectors like the CDC has to be taken into account. For 85\% of the cluster on one side was one partner on the other side. This number reflects the detector and readout efficiency.   
\subsection{Trigger Setup}
\subsubsection{Trigger Settings}
During the p+p data taking with the $\mathrm{\Lambda}$ Trigger installed within the FOPI Spectrometer three different reaction triggers were used.
\begin{itemize}
    \item \textbf{LVL1:} The standard FOPI first level trigger was derived from the particle multiplicities in the time of flight detector of the FOPI Spectrometer. The minimal condition was a multiplicity of one hit in the forward ($4\degree\le \theta \le 28\degree$) and one in the backward ($32\degree\le \theta \le 110\degree$) directions.
    \item \textbf{\silvioo:} The \silvio trigger condition required that the number of hits on the \silvioo-A layer ($\mathrm{M_{A}}$) and on the \silvioo-B layer ($\mathrm{M_{B}}$) was set to the conditions $\mathrm{0<M_{A}}$ and $\mathrm{1<M_{B}}$.
    \item \textbf{LVL2:} The second level trigger conditions results from the logical AND combination of the main first level and the \silvio trigger condition.\\
\end{itemize}
\subsubsection{Multiplicity Selection}
\begin{figure}[ht]
  \begin{picture}(252,128)
    \centering    
    \put(0,0){\includegraphics[width=0.48\columnwidth]{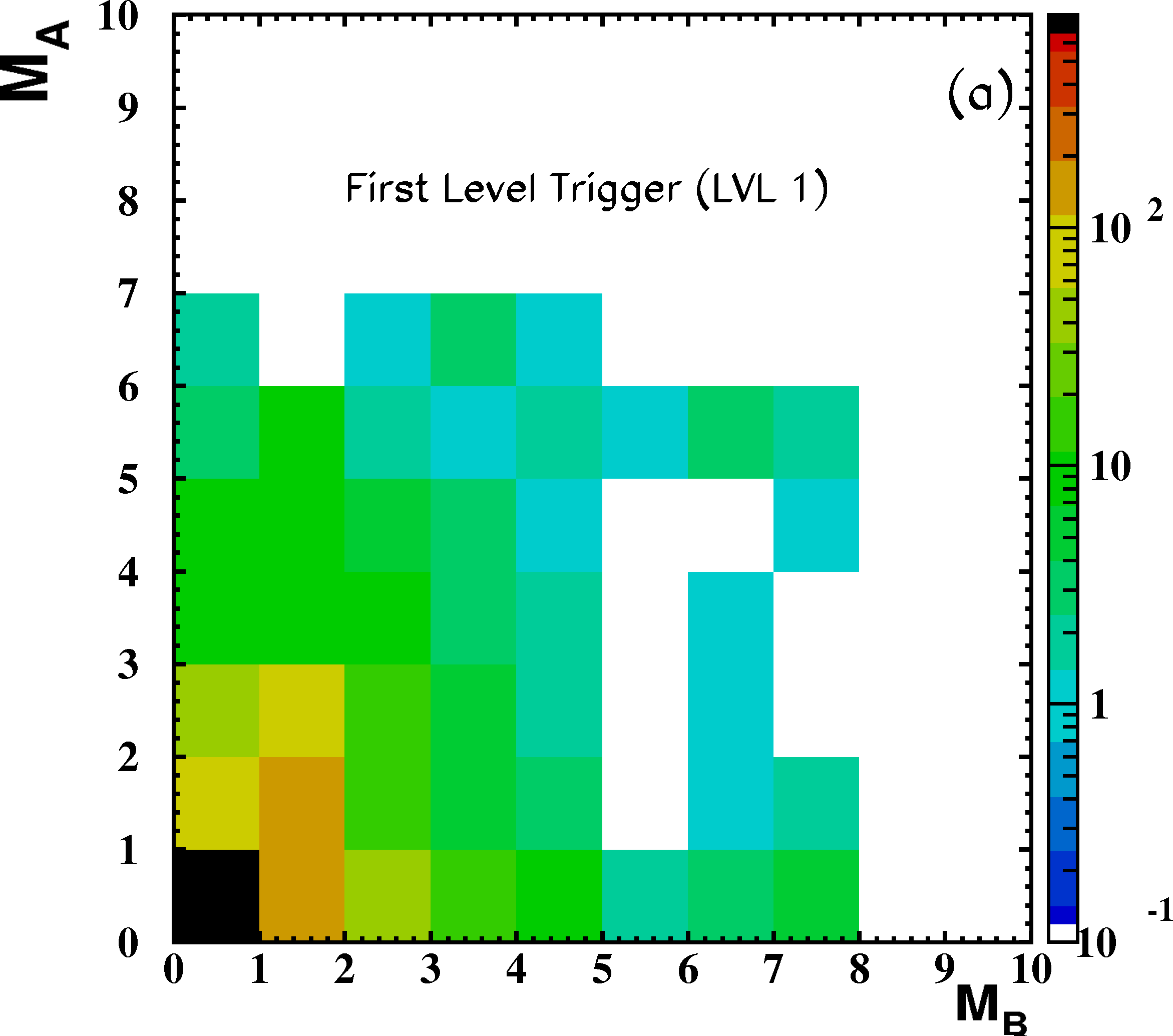}}
    \put(128,0){\includegraphics[width=0.48\columnwidth]{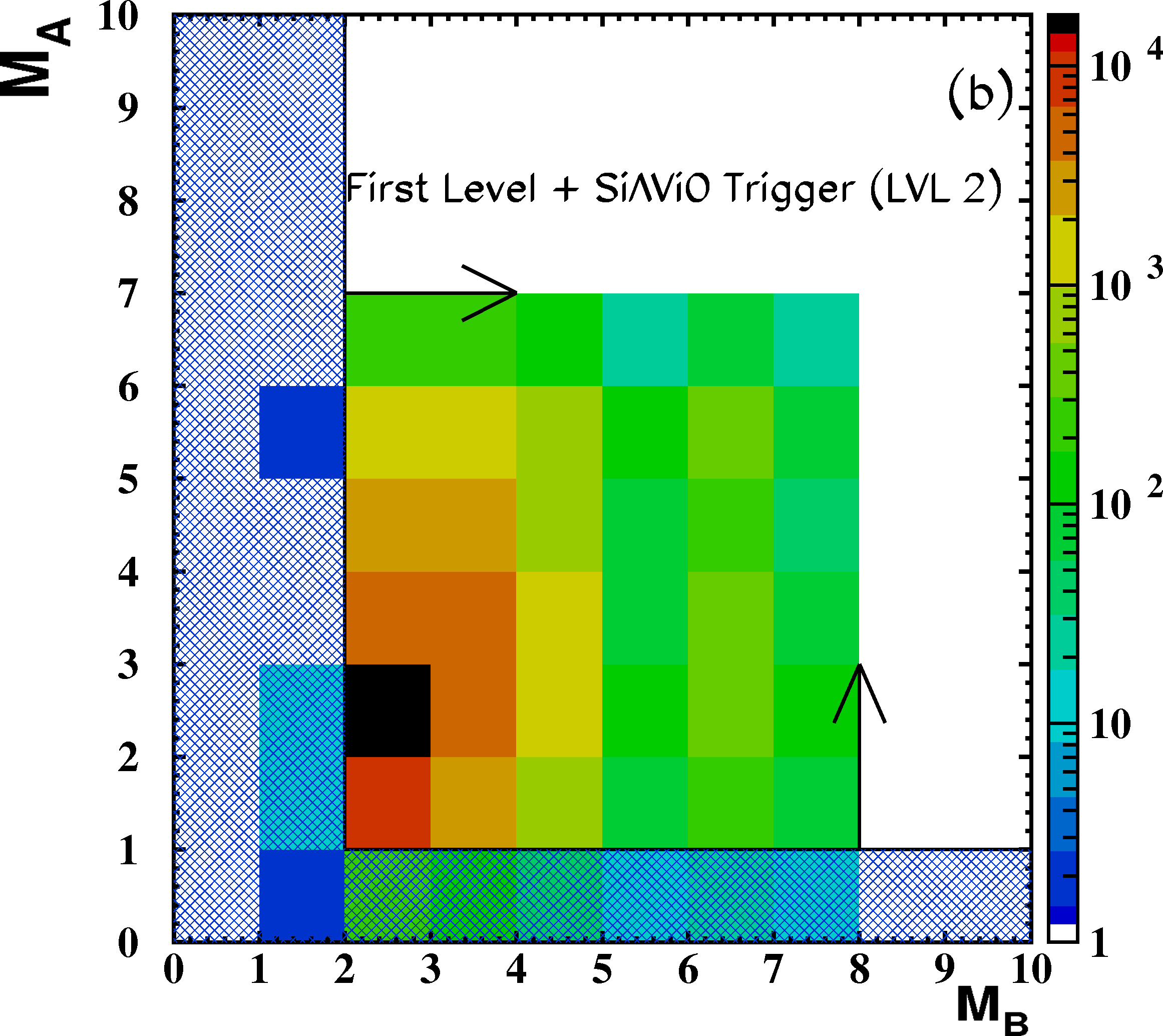}}
  \end{picture}  
    \caption{Hit-Multiplicity of \silvioo-A layer versus \silvioo-B layer. Picture (a) shows the distribution of first-level trigger event; picture (b) for second level triggers. The blue shaded area correspond to fake events.}
  \label{pic:soft:mult}
\end{figure}

The effect of the second level trigger on proton-proton reaction data is shown in figure~\ref{pic:soft:mult}, in which the recorded hit multiplicity on the \silvio A versus the multiplicity on \silvio B is displayed.
One can see the effect of the multiplicity selection obtained by applying the \silvio trigger condition (b) in addition to the first level condition (a). The amount of events with a wrong multiplicity condition ((b) blue shaded area) is below 0.7\%.
\subsubsection{Trigger Rates}
\label{sec:triggerrates}
Since the \silvio detector is located directly behind the target, the \silvio trigger increases also the amount of events originating from the target.
To determine the fraction of trigger reactions, which are not coming from the target, 
the reaction ratio of the first and second level trigger were measured with and without a filled liquid hydrogen target.
The resulting ratios are shown in table~\ref{tab:triggerratio} together with the first and second level trigger rates.
For the target measurement the reduction from LVL1 to LVL2 is in the order of a factor 10 while for the case without target a reduction of a factor of about 100 was measured.
Furthermore a comparison between the two cases for each trigger condition shows that the LVL1 trigger is not selective for target events while for the LVL2 condition
a significant difference between the target and the no target case can be seen.
This clearly indicates that the \silvio trigger condition that determines the most restrictive condition within the LVL2 trigger corresponds to a target-trigger as well.
Furthermore if one takes the empty-target data and the corresponding LVL2 trigger rate as a baseline, one can extract a target event purity of the LVL2 trigger to be a the level of $\approx$90\%.
This shows that FOPI without the \silvio trigger is not well suited for triggering on elementary reactions and the usage of the \silvio trigger is mandatory.
\begin{table}[ht]
\centering
  \begin{tabular}{|c|c|c|}
    \hline
    Trigger per Beam particle & LVL1               & LVL2               \\\hline
    Target    & $ 3.1\pm0.1\cdot10^{-3}$ & $ 3.8\pm 0.1\cdot10^{-4}$\\\hline
    No target & $ 2.6\pm0.1\cdot10^{-3}$ & $ 3.9\pm 0.1\cdot10^{-5}$\\\hline
    Ratio     & $80.9\pm1.3$\%           & $10.4\pm 1.1$ \%         \\\hline
  \end{tabular}
  \caption{Amount of trigger per beam particle for LVL1 and LVL2 trigger. The last row shows the ratio of target and no target rate.}
  \label{tab:triggerratio}
\end{table} 
\section{Tracking and $\Lambda$ reconstruction}
\subsection{Matching and Refitting}
\begin{figure}[ht]
  \centering
  \includegraphics[width=.8\columnwidth]{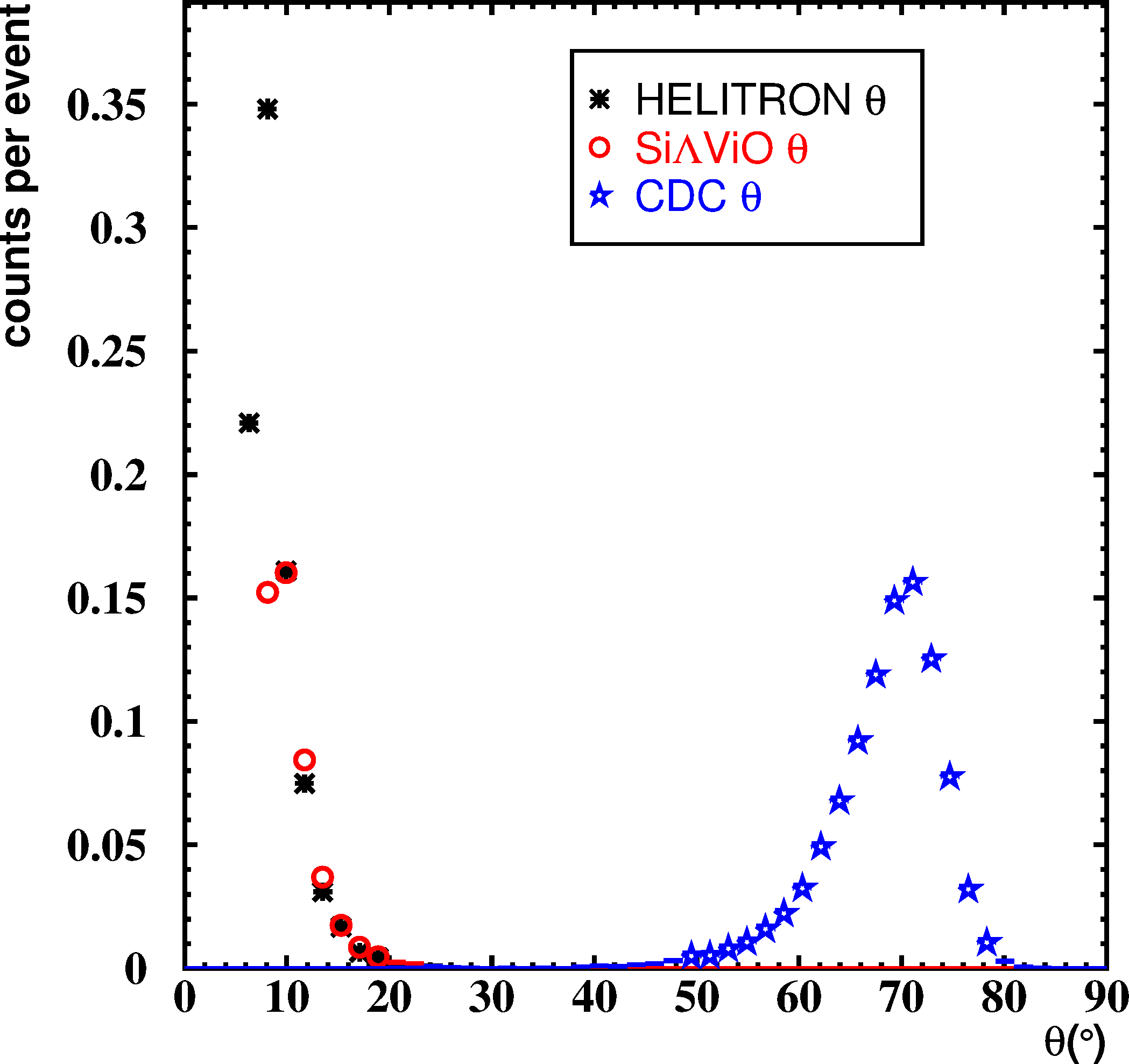}
  \caption{Polar angle distributions of reconstructed tracks in the \silvio Detector (red circles) and the drift chambers Helitron (black crosses) and CDC (blue stars) of elastic proton-proton reactions.}
  \label{pic:soft:momrefit}
\end{figure}

Since \silvio is meant to improve the tracking performances in the forward direction, it was build to overlap geometrically with the forward drift chamber Helitron. This can be seen quite good in the distribution of the polar angle of reconstructed tracks in \silvio and the two drift chamber CDC and Heltiron (figure~\ref{pic:soft:momrefit}) for elastic proton-proton reactions. In the forward direction \silvio totally overlaps with the Helitron drift chamber between 10\degree\ and 20\degree. Below 10\degree\  the acceptance of \silvio is lower, due to the shape of the \silvio B plane. The hit points of the \silvio detector have first to be combined with the information delivered by the other FOPI sub-detectors and then in a second step the tracks have to be refitted.

\subsubsection{Matching}    
\begin{figure}[ht]
  \begin{picture}(252,128)
   \put(0,0){\includegraphics[width=.48\columnwidth]{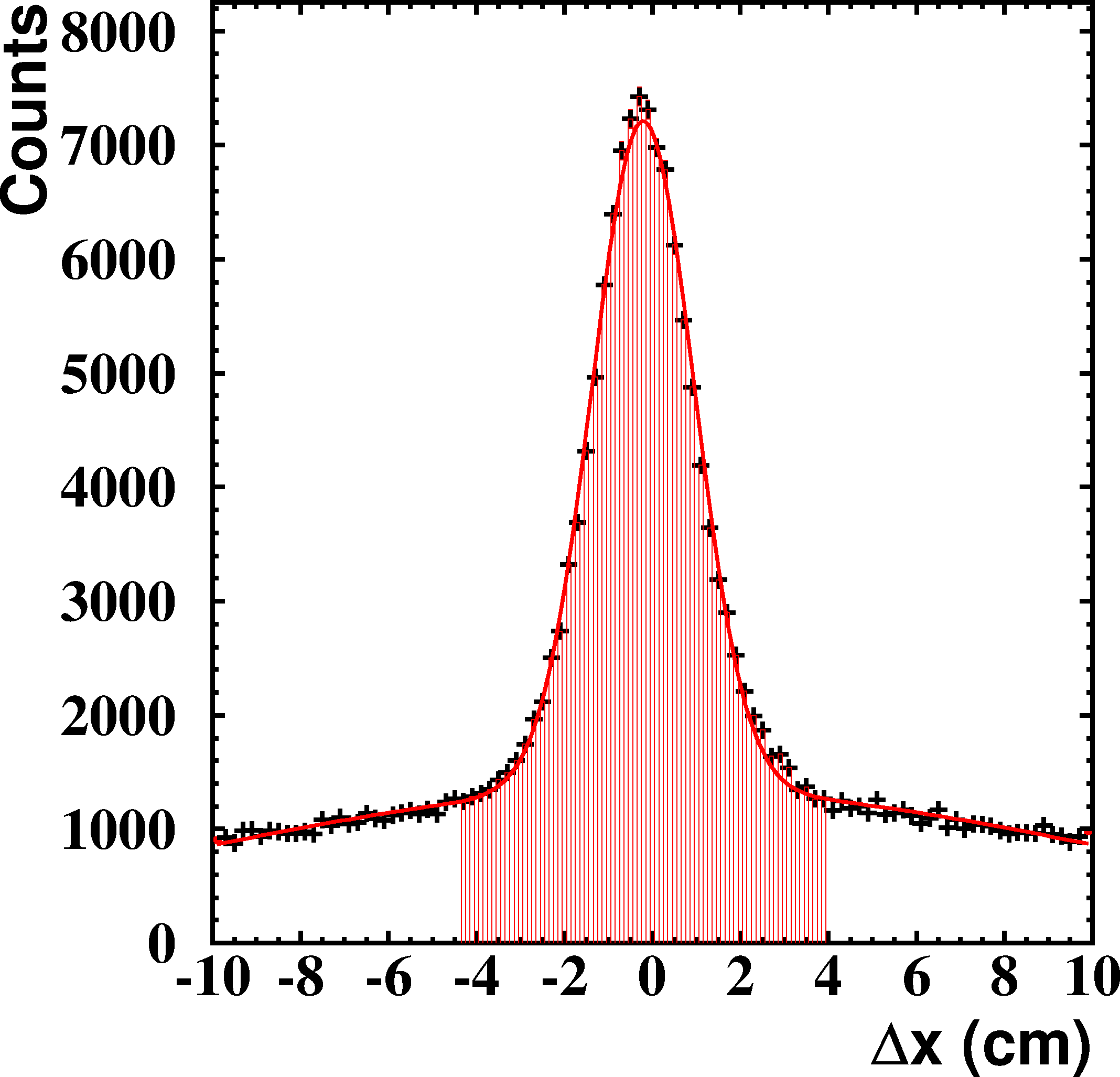}}
   \put(30,100){(a)}
   \put(75,105){$\mu=-0.21$}
   \put(75,95){$\sigma=\,\,\,1.15$}
   \put(136,0){\includegraphics[width=.48\columnwidth]{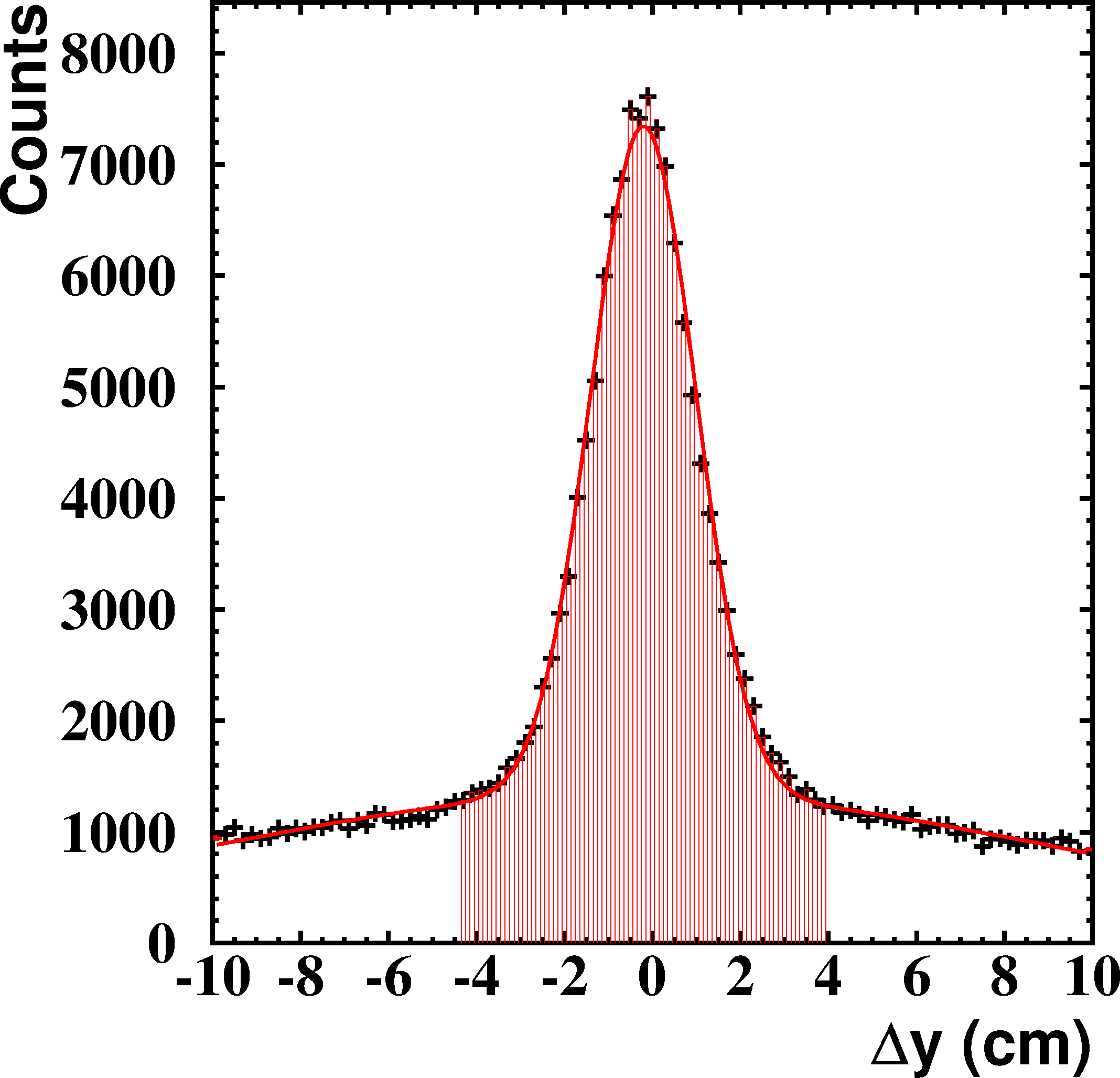}}
    \put(166,100){(b)}
    \put(211,105){$\mu=-0.20$}
    \put(211,95){$\sigma=\,\,\,1.18$}
   \end{picture}
   \caption{Distance between the \silvio B hit point and the extrapolated point from the Helitron track to the plane perpendicular to the beam axis and located at the same z position as the \silvioo-B  layer. Panel (a) and  (b) shows the x and y coordinates of the distance vector respectively.}
   \label{pic:soft:matchdist}
 \end{figure}
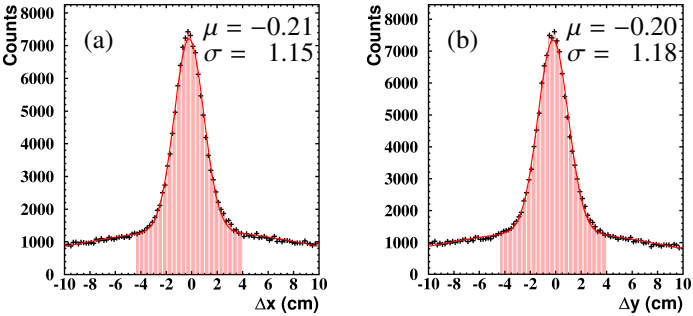

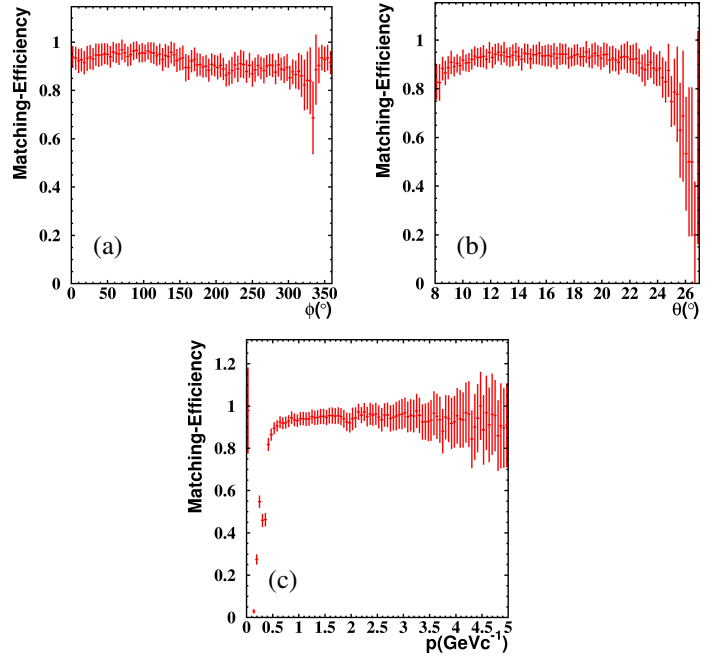
\begin{figure}[ht]
    \begin{picture}(252,252)
    \put(0,126){\includegraphics[width=.48\columnwidth]{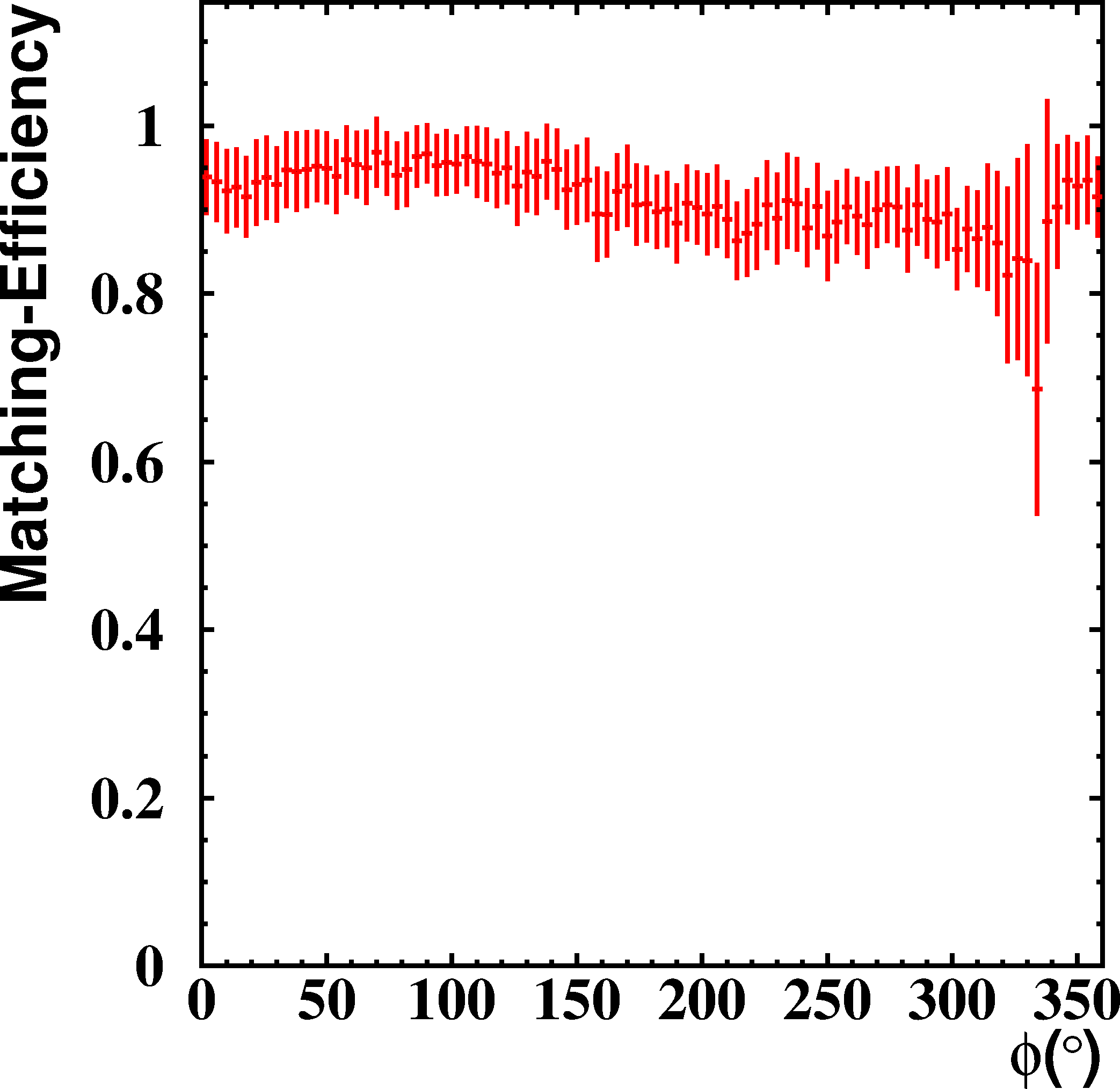}}
    \put(30,150){(a)}
    \put(136,126){\includegraphics[width=.48\columnwidth]{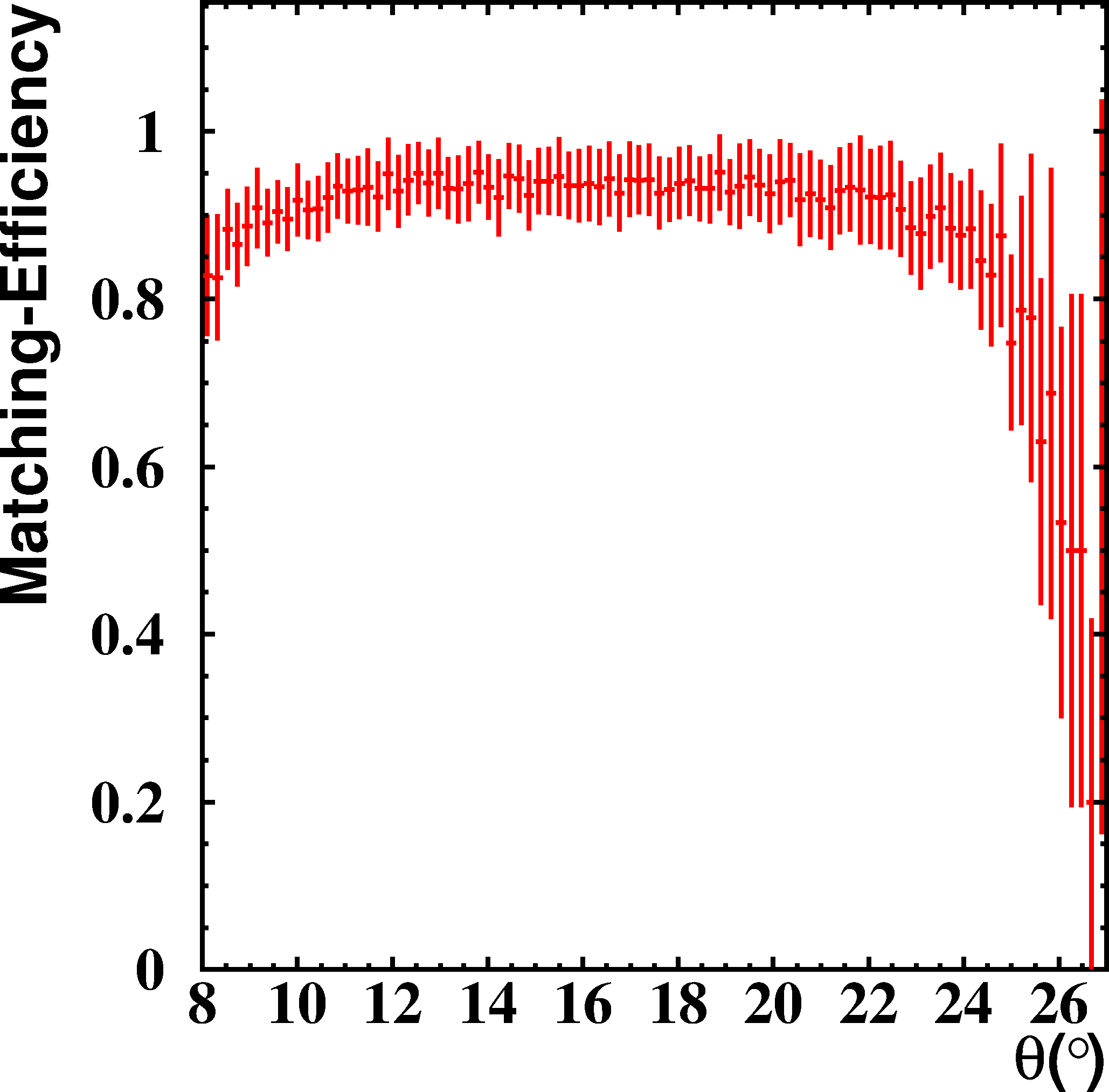}}
    \put(166,150){(b)}
    \put(65.52,0){\includegraphics[width=.48\columnwidth]{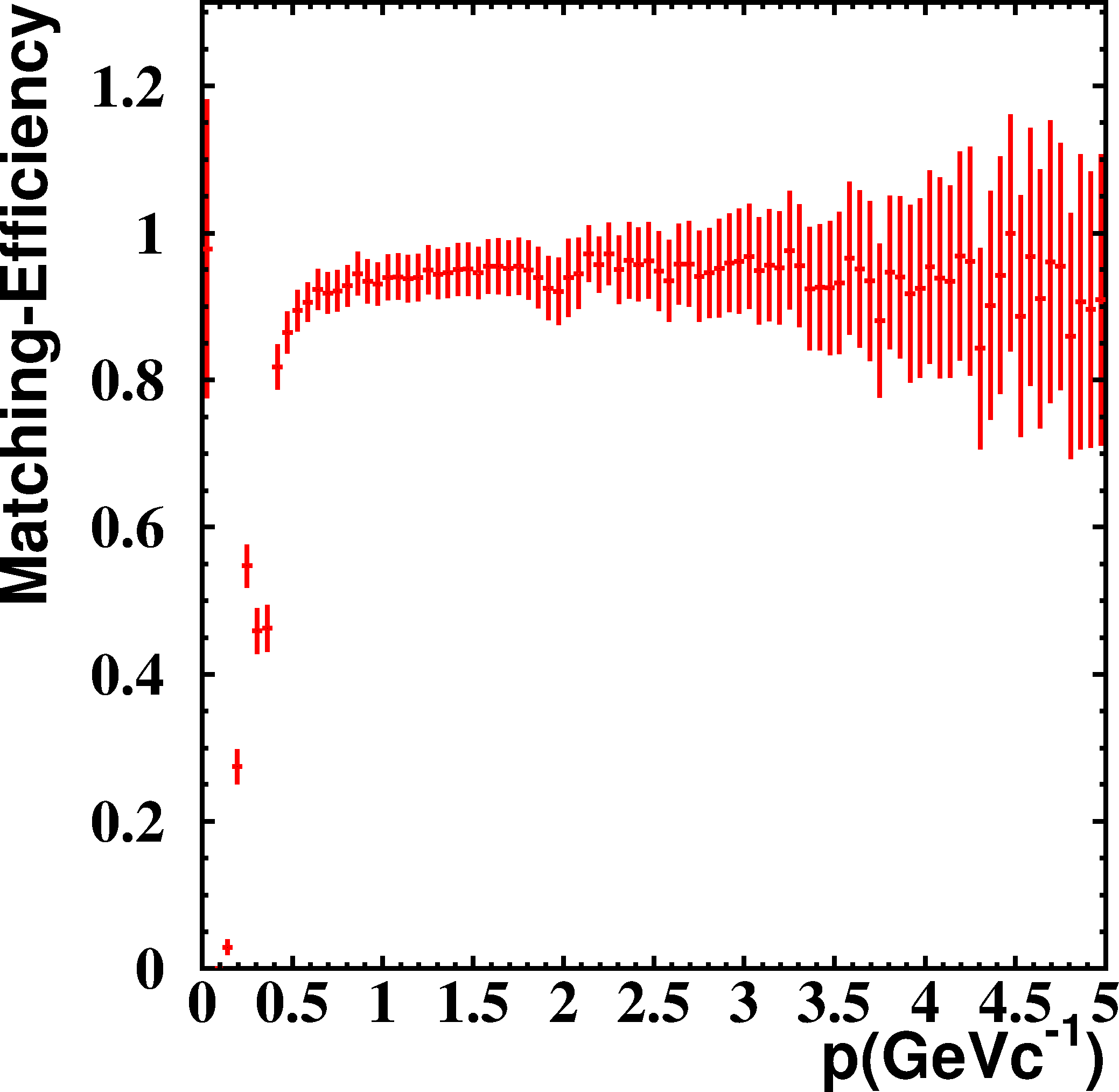}}
    \put(95.52,24){(c)}
    \end{picture}
  \caption{Matching efficiency of \silvio hit points with Helitron tracks within the \silvio acceptance versus the azimuthal angle $\mathrm{\phi}$ (a), the  polar angle $\mathrm{\theta}$ (b) and the particle momentum.}
  \label{pic:soft:matcheff}
\end{figure}

Each hit point on the \silvioo-B layer is combined with the track information of the Helitron drift chamber of FOPI via geometrical matching. 
Since the particles are flying through the solenoid magnetic field of FOPI, the tracks of the drift chambers are extrapolated up to the plane orthogonal to the beam axis at the same z-position as the \silvioo-B layer. 
Figure~\ref{pic:soft:matchdist} shows the distance between the reconstructed position on the \silvioo-B  detector and the extrapolated point from the tracks in the forward drift chamber. The width of the distribution is mainly caused by the measurement of the Helitron.\\
All combinations within a $\mathrm{5\sigma}$ cut on the distance for both the x and y coordinates are accepted. In case of more than one accepted \silvioo-B hit, the hit with the smaller distance is chosen.\\
The amount of mismatched tracks was determined by simulation and found to be lower than 5\%. \\
The ratio of Helitron tracks matched with \silvio hits to all Helitron tracks falling within the \silvio acceptance, which corresponds to the matching efficiency, is shown in figure~\ref{pic:soft:matcheff} as a function of different track parameters: the laboratory angles $\mathrm{\theta}$ and $\mathrm{\phi}$ and the momentum $\mathrm{p}$. One can see a homogeneous distribution with an average matching efficiency of 91.2$\pm$ 0.6 \%.

\subsubsection{Refitting - Momentum Resolution Improvement}
By adding the \silvio hit point to the track reconstruction in the forward direction one can improve the momentum resolution.
The track fitting is repeated considering this additional point, which can improve the momentum resolution slightly, which is dominated by the resolution of around 8 cm along the drift wires in the Helitron (perpendicular to the beam axis) \cite{plettner}.\\
In the first step the polar angle of the particle is recalculated.
Since FOPI is using a solenoid magnetic field, the particle trajectories are straight in the rz-plane.
Without \silvioo, the polar angle of the particle is calculated using the hit-point information of the plastic wall, under the assumption, that the particle originated from the vertex.
With the \silvio hit point the vertex-assumption can be discarded and a more realistic theta can be calculated. 
The second step is to fit a track to all the available hit points and calculate the momentum of the particle \cite{wind}.
The resolution improvement, achievable with this procedure, was tested with the help of simulations.
The resolution is defined as:
\begin{equation}
\text{Resolution}=\frac{p_{reco}-p_{sim}}{p_{sim}}=\frac{\Delta p}{p}
\label{def:momres}
\end{equation}
\begin{figure}[H]
  \begin{picture}(252,128)
    \put(0,0){\includegraphics[width=0.48\columnwidth]{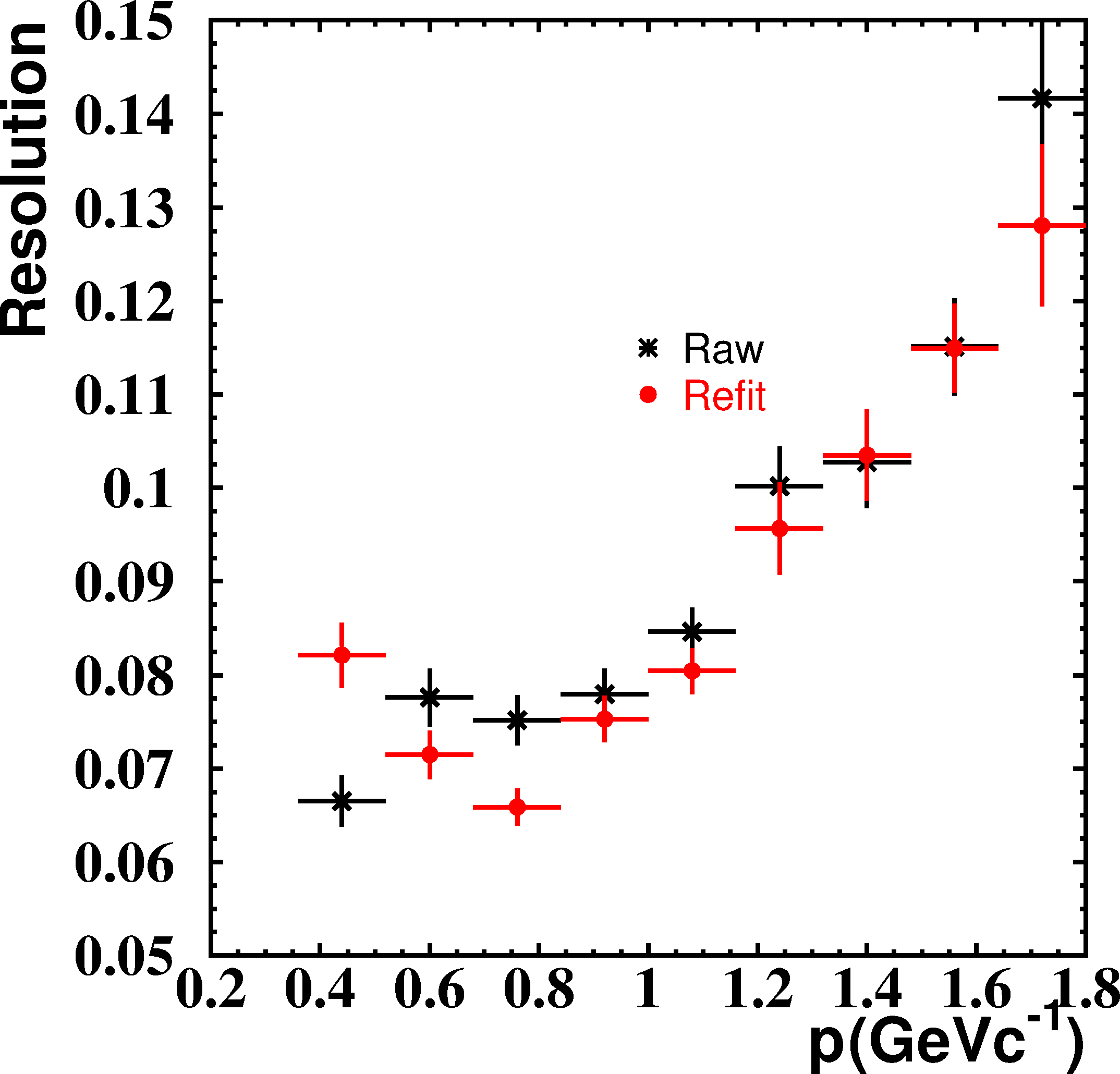}}
    \put(30,100){(a)}
    \put(136,0){\includegraphics[width=.48\columnwidth]{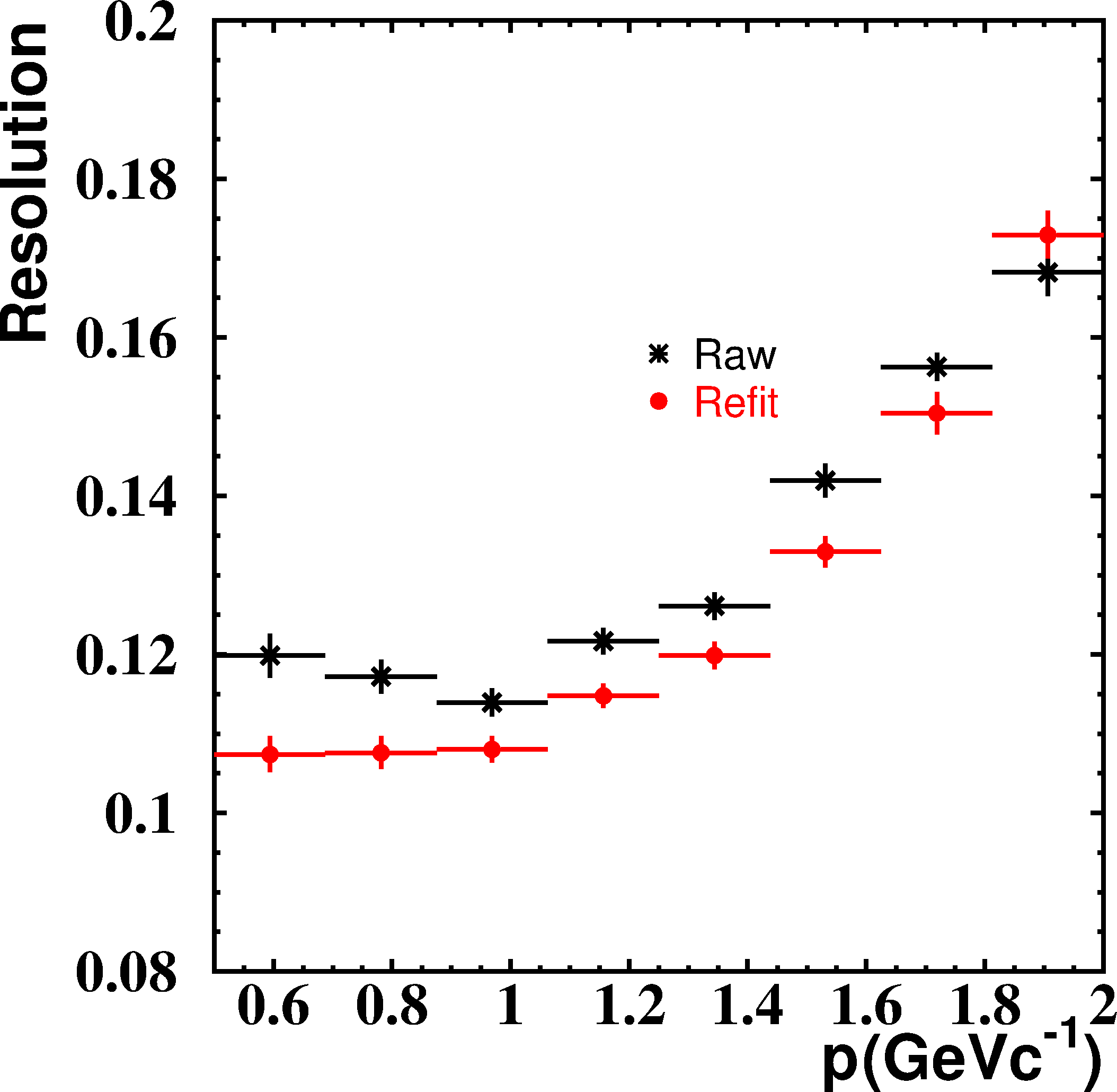}}
    \put(166,100){(b)}
  \end{picture}
  \caption{The left row shows the momentum resolution for pions (panel (a)) and protons (panel (b)), for Helitron alone (black) and after refitting with the \silvio point (red) for the second level trigger condition.}

\label{pic:soft:pipres}
\end{figure}
The resolution for pions and protons with and without the refit is shown in figure~\ref{pic:soft:pipres}, as the width of the distribution of equation~\ref{def:momres} for the second level trigger condition.
One observe that for the pions an average improvement of 4\% is possible while for the protons more than 5\% improvement is achieved. The limited improvement  is caused, by the fact, that the momentum resolution is manly dominated by the point resolution in the Helitron.

\subsection{$\mathrm{\Lambda}$ Offline Reconstruction}
\label{soft:lam:rec}
\begin{figure}[H]
  \centering
  \includegraphics[width=.8\columnwidth]{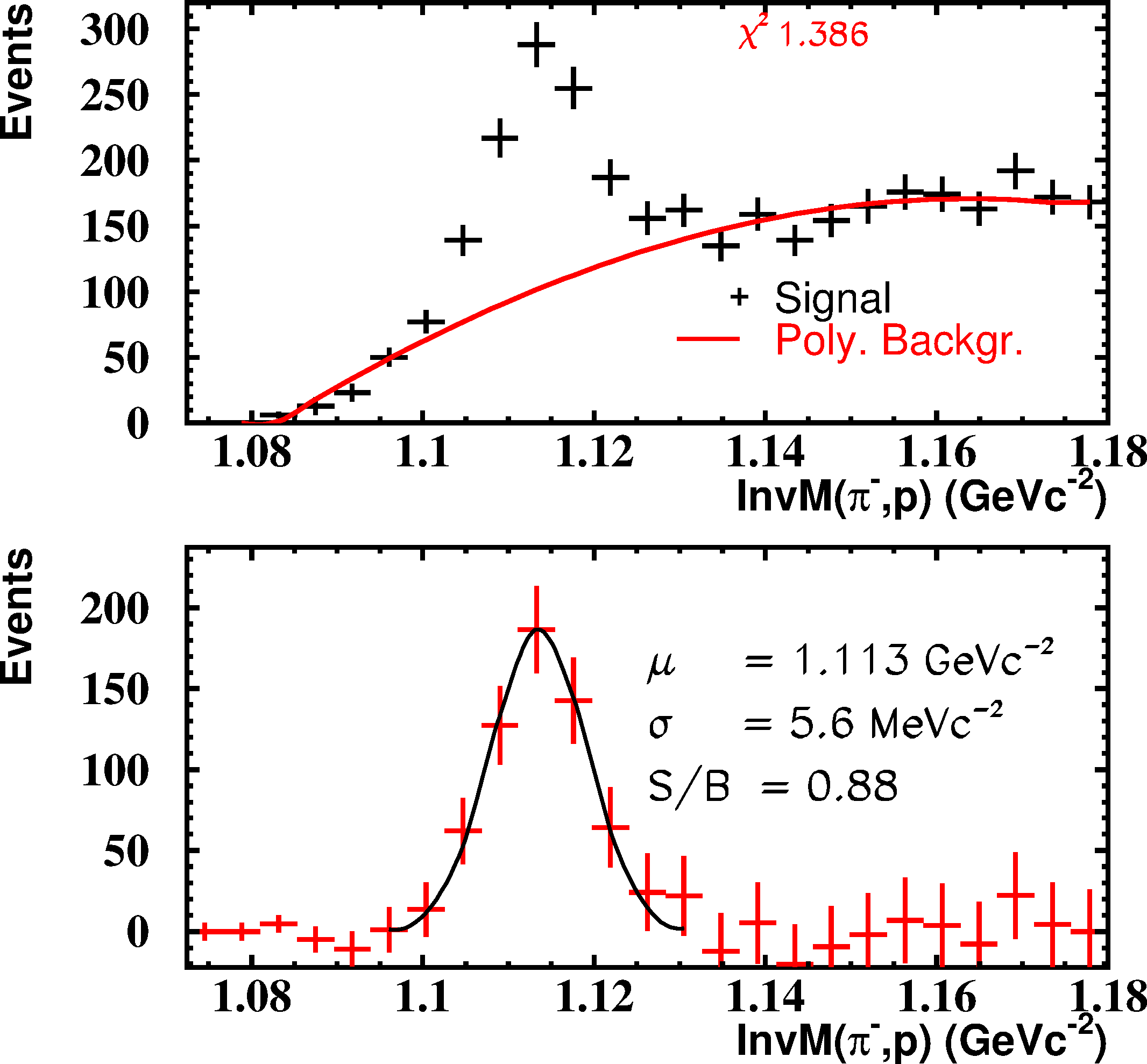}
  \caption{Invariant mass of the $\mathrm{p-\pi^-}$ ($\mathrm{\pi^{-}}$ measured in CDC, p in Helitron) pairs from UrQMD simulation of the p+p at 3.1 GeV reaction. The red line in the upper panel shows the polynomial fit to the combinatorial background. The lower panel shows the background-subtracted spectrum. The black line corresponds to a Gaussian fit to the signal. }
  \label{pic:invLSimu}
\end{figure}

\begin{figure}[H]
  \centering
  \includegraphics[width=.8\columnwidth]{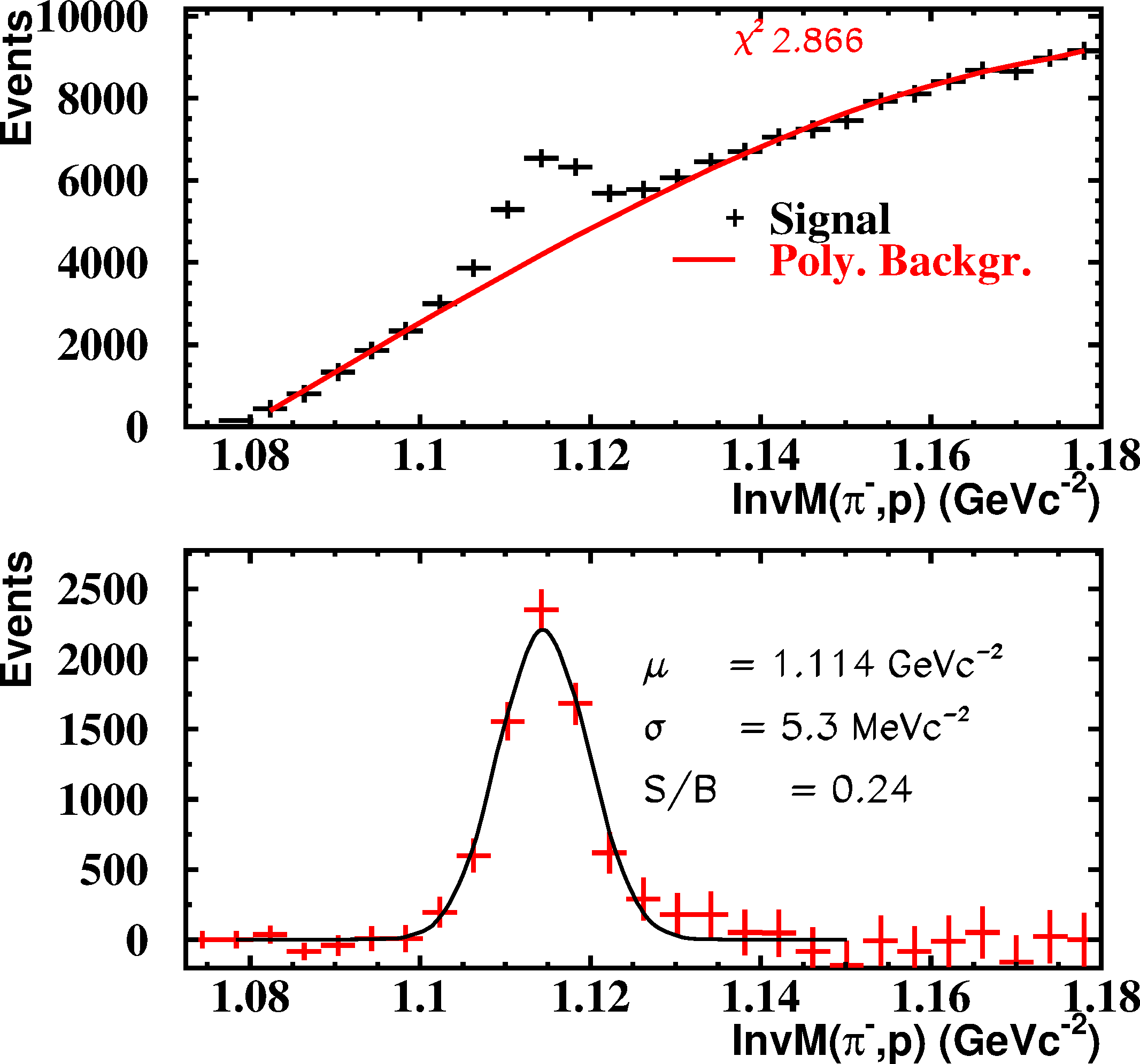}
  \caption{Invariant mass of the $\mathrm{p-\pi^-}$ ($\mathrm{\pi^{-}}$ measured in CDC, p in Helitron) pairs from experimental data. The red line in the upper panel shows the polynomial fit to the combinatorial background. The lower panel shows the background-subtracted spectrum. The black line corresponds to a Gaussian fit to the signal.}
  \label{pic:invLData}
\end{figure}

$\mathrm{\Lambda}$ hyperons are reconstructed via their decay into a negative pion and a proton. The invariant mass of all proton and pion pairs either indentified in the CDC or the Helitron is calculated. The usual particle identification cuts employed within the FOPI analysis have been used \cite{FOPILAM2}.\\
To reduce the amount of uncorrelated background additional cuts on the decay vertex have to be applied. The presence of \silvio allows to calculate a vertex position in case a particle is flying though the Helitron, which would not be possible without the additional track point, due to the vertex assumption of the tracking procedure. Since the $\mathrm{\Lambda}$ hyperon has a decay length $\mathrm{c\tau=}7.86$ $\mathrm{cm}$ on average the vertex of its decay position can be separated from the primary vertex. Other uncorrelated proton-pion pairs are stemming from the primary vertex.\\
In order to select $\mathrm{\Lambda}$ hyperon candidates, the position of the primary vertex is recalculated under the assumption that each $\mathrm{p-\pi^-}$ combination stems from a $\mathrm{\Lambda}$ decay. The constructed '$\mathrm{\Lambda}$' vector\footnote{'$\mathrm{\Lambda}$' vector: Sum of proton and $\pi^{-}$ momentum} is used together with the other measured particle vectors to determine the primary vertex position. If more than one combination is found, these for which the recalculated vertex position is outside the target volume is discarded. These specific cuts applied to the recalculated primary vertex position $primvertex_{x,y,z}$ are summarized in table~\ref{tab:cut}. As additional cut on the distance between the primary and the secondary vertex $\mathrm{dr}$ is applied to reduce the combinatorial background. All cut values used are listed in table~\ref{tab:cut}.
Figure~\ref{pic:invLSimu} and figure~\ref{pic:invLData} show reconstructed $\mathrm{\Lambda}$ particles for UrQMD simulation and experimental data for the reaction p+p at 3.1\,GeV.
The upper panels show the invariant mass of all p-$\mathrm{\pi^{-}}$ pairs after applying  the vertex cuts mentioned above.
These spectra were fitted by the sum of a Gaussian and a polynomial function. The polynomial function is represented by the red curve.
The lower panels show the background-subtracted signal of the invariant mass.
The black line represents a Gaussian fit to this spectrum and delivers the mean ($\mathrm{\mu}$) and the width ($\mathrm{\sigma}$) of the reconstructed $\mathrm{\Lambda}$ particles. One can see that a mass resolution (expressed in $\mathrm{\sigma}$) of $\mathrm{5.6\,MeV/c^2}$ and $\mathrm{5.3\,MeV/c^2}$ has been achieved for the simulated and experimental data respectively. These results refer to $\mathrm{\Lambda}$ candidates where the $\mathrm{\pi^-}$ candidate was detected in the CDC and the proton in the Helitron.One can see, that the data and simulation - including the trigger selection - are rather comparable as far as the resolution is concerned but the $\mathrm{\Lambda}$ yield looks different, as well as the signal to background ratios of 0.88 in simulation  and 0.24 in data. The optimization procedure of the detector digitizers will not be discussed in this work, but the efficiency of the apparatus is rather well under control in simulations. Still the $\mathrm{\Lambda}$ inclusive production cross-section and the simulated angular distribution are directly coming from UrQMD and could be rather far away from the experimental data. A dedicated tuning of these variables is currently on-going to finalize the physics results extracted from this measurement.
\begin{table}[H]
\centering
  \begin{tabular}{|c|c|c|}            
    \hline
    value & minimal value [cm] & maximal value [cm] \\\hline
    $\mathrm{primvertex_{x}}$ & -1.0 &  1.0 \\
    $\mathrm{primvertex_{y}}$ & -1.0  & 1.0 \\    
    $\mathrm{primvertex_{z}}$ & -2.0  & 2.0 \\
    $\mathrm{dr}$ & 3.0  & $\infty$ \\    
    \hline
  \end{tabular}
  \caption{Cut values used for the extraction of the $\mathrm{\Lambda}$ signal shown in figures~\ref{pic:invLSimu},~\ref{pic:invLData}. }
  \label{tab:cut}
\end{table}
\section{Trigger Performance}
\subsection{Event Reduction and Background Suppression}
\label{sec:signal:back}
The trigger settings described in section~\ref{TrigPer} are meant to increase the $\mathrm{\Lambda}$ content of the recorded events but they obviously introduce a bias for the accepted events. This effect must be accounted for if one wants to extract final cross-sections and can be estimated with the help of simulations by comparing reconstructed data to Monte Carlo tracks.
For this purpose the $\mathrm{pK\Lambda}$ signal was generated with PLUTO \cite{Pluto} programm including the angular distribution extracted from \cite{cosy3} for the $\mathrm{\Lambda}$ hyperon. For the simulation of the background UrQMD events without strangeness production \cite{URQMD} were used. \\
\begin{table*}[ht]
\centering
\begin{tabular}{|c|c|c|}
\hline
           & LVL1   & LVL2   \\
\hline
Signal ($\mathrm{pK\Lambda}$) & $0.52 \pm 0.003 (stat) \pm 0.16 (ang) $ & $0.29 \pm 0.002 (stat) \pm 0.09 (ang) $ \\
Background & $0.183 \pm 0.003$ & $2.51 \pm 0.02 \cdot 10^{-2}$ \\
\hline
$\mathrm{pK\Lambda}$ Bkg. Supp. & $2.86 \pm 0.05 (stat) \pm 0.91 (ang)$ & $11.93 \pm 0.12 (stat) \pm 3.7 (ang)  $ \\
\hline
\end{tabular}
\caption{Events acceptance for signal and background events for different trigger conditions. For the signal acceptance an angular distribution taken from \cite{cosy3} was taken into account. The resulting error values for the angular correction (ang) are shown separated from the statistical errors (stat).}
\label{sig:tab:red}
\end{table*}
In table~\ref{sig:tab:red} the acceptance for the signal and background events are listed for different trigger conditions as well as the resulting background reduction for the LVL1 and LVL2 condition. The error values for the background suppression are dominated by the error values of the angle correction taken from \cite{cosy3}.\\
The rather low acceptance for the signal events is mainly caused by the fact that a charged particle hit is required in the region covered by the RPC or Barrel detector. Indeed a good kaon identification is mandatory to select the reaction $\mathrm{p+p}$ $\mathrm{\rightarrow p+K^++\Lambda}$ and this is possible only exploiting the RPC device. This explain the numbers shown in table~\ref{sig:tab:red}.\\ 
For the LVL1 the background reduction is $2.86 \pm 0.05 (stat) \pm 0.91 (ang)$ and for the the LVL2 trigger condition $11.93 \pm 0.12 (stat) \pm 3.7 (ang)$, which corresponds to a signal to background from LVL1 to LVL2 within the geometrical acceptance region of a factor $4.09 \pm 0.08 (stat) \pm 1.8 (ang)$. The statistical (stat) errors and the ones from the angular correction (\cite{cosy3}) (ang) are shown separately. Moreover the LVL2 settings enable an efficient suppression of the off-target events, as discussed in section~\ref{sec:triggerrates}. The efficiency of such selection yields about $90\%$, which translates into $10\%$ of events that stem from off-target reactions and are recorded by the LVL2 trigger. On the other hand the LVL1 sample is highly contaminated by off-vertex events and this contamination, which was evaluated quantitatively from the empty-target measurement, must be accounted for in the evaluation of the trigger performance.
\subsection{Enhancement of $\mathrm{\Lambda}$ Events }
\label{sec:res:lamenh}
\begin{figure}[ht] 
 \begin{picture}(252,128)
   \centering
   \put(0,0){\includegraphics[width=.48\columnwidth]{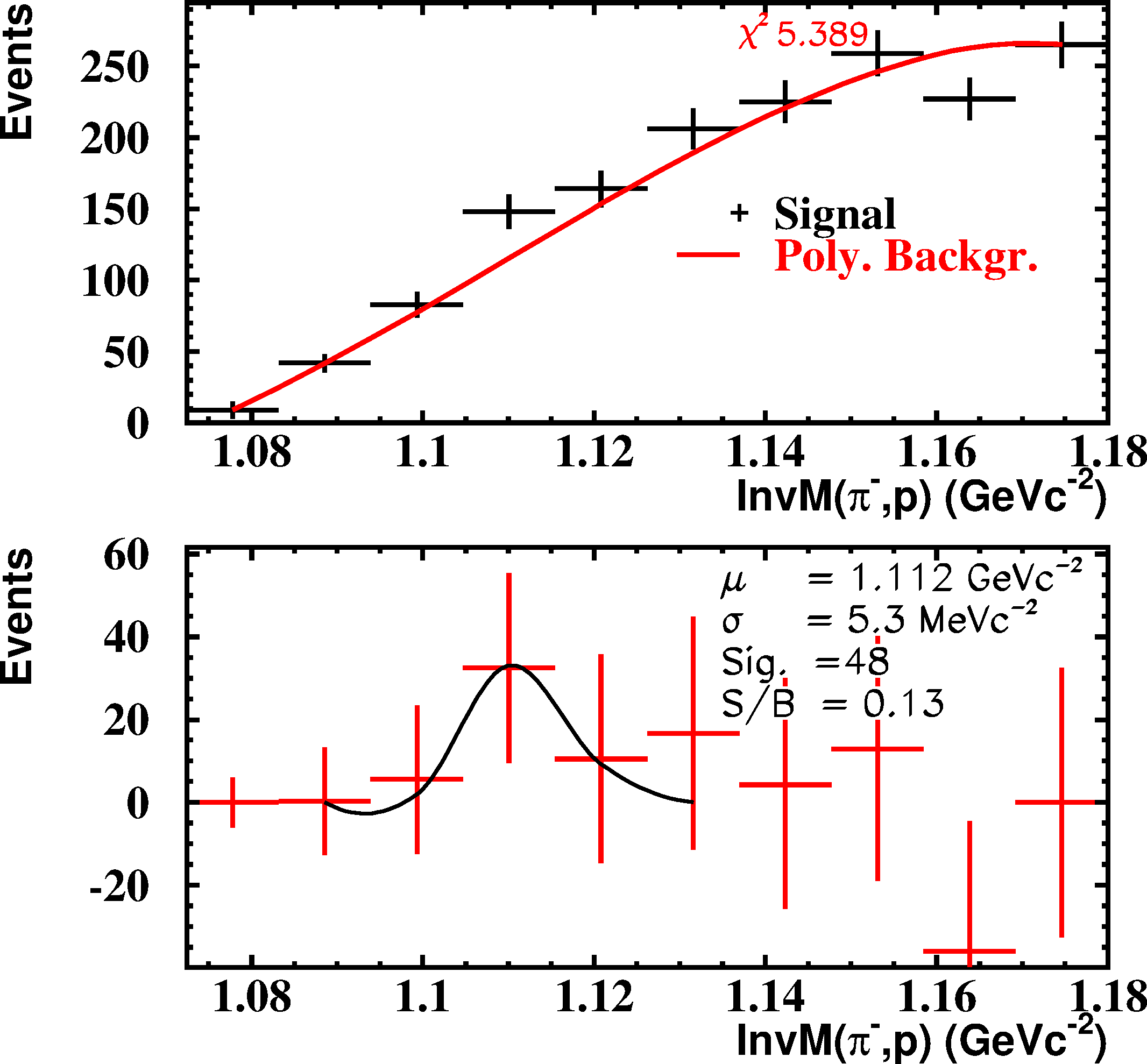}}
   \put(30,100){(a)}
   \put(136,0){\includegraphics[width=.48\columnwidth]{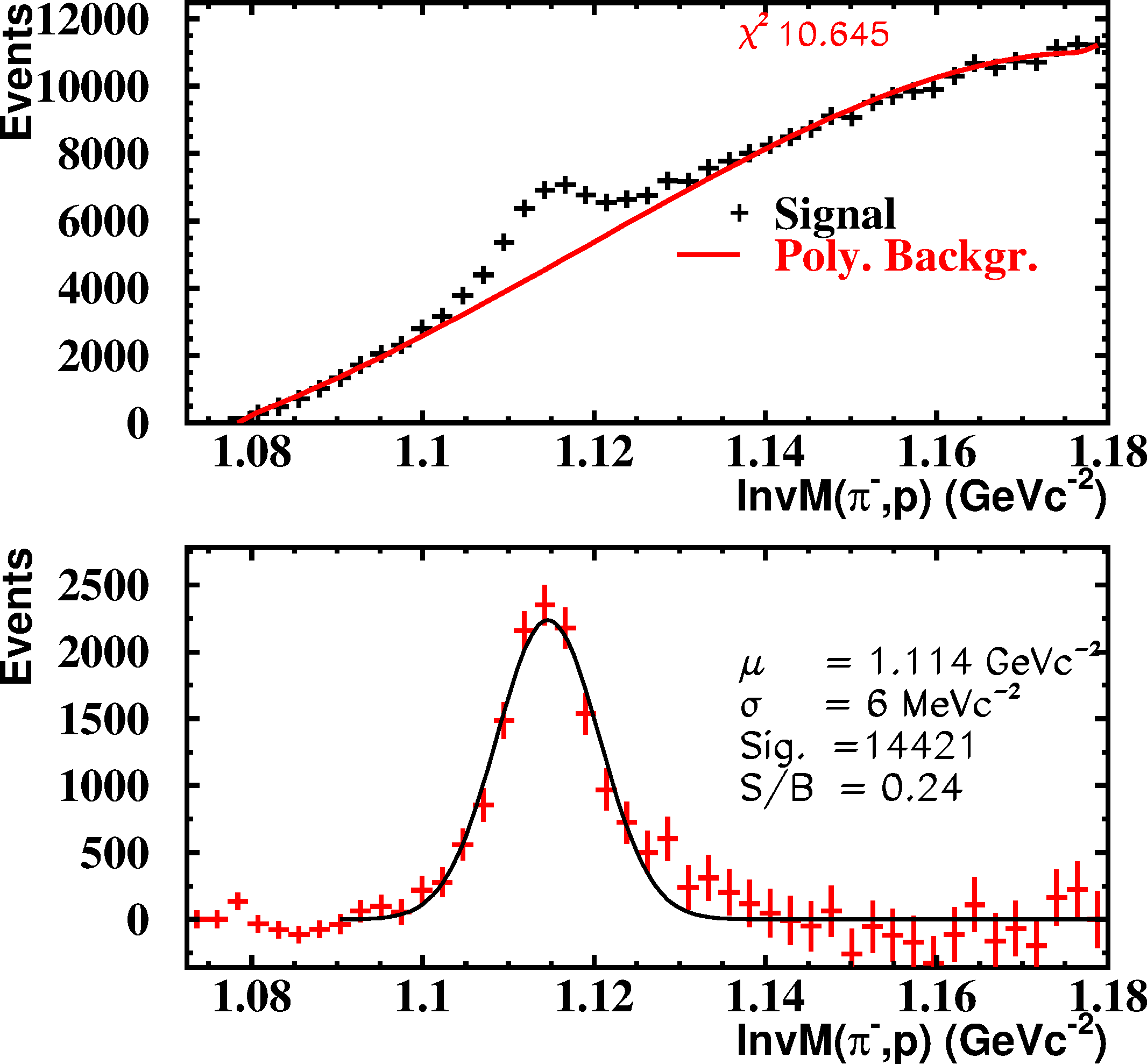}}
    \put(166,100){(b)}
 \end{picture}
   \caption{Invariant mass of proton and pion for $\mathrm{pK\Lambda}$ LVL1 (a) and LVL2 (b) trigger condition. The black crosses in the upper picture show the signal and the red line shows the fitted polynomial background. The lower pictures show the background subtracted signal.}
\label{pic:results:invltrigger}
\end{figure}

To determine the experimental enhancement of the $\mathrm{\Lambda}$ signal reached through the LVL2 settings, the reconstructed $\mathrm{\Lambda}$ signal per event for the LVL2 and the LVL1 trigger conditions was compared. For this purpose, $\mathrm{\Lambda}$-hyperons were reconstructed in both data samples, employing a wider selection cuts in dr ($\mathrm{dr} > 1.0$ cm) than in table~\ref{soft:lam:rec} in order to enhance the signal content; again $\mathrm{\pi^{-}}$ in the CDC and proton in the Helitron hemisphere were selected.
Figure~\ref{pic:results:invltrigger} shows the resulting invariant mass plot distributions obtained for the LVL1 (left panel) and LVL2 (right panel) samples. Due to the less stringent cut conditions, the width of signal increased slightly with respect to the results discussed in section~\ref{soft:lam:rec} from  $\mathrm{5.3\,MeV/c^2}$ to $\mathrm{6.0\,MeV/c^2}$, while the signal to background ratio stays quite stable at a value of $0.24$. The signal yield was extracted by integrating the distribution in a $\mathrm{3\sigma}$ interval around the fitted mean value $\mathrm{\mu}$.
One can see that the $\mathrm{\Lambda}$ signal is barely visible for the LVL1 events, even if figure~\ref{pic:results:invltrigger} corresponds to the total collected statistics;  indeed the signal to background for the reconstructed $\mathrm{\Lambda}$ goes down to 0.13. For this reason a lower limit of the $\mathrm{\Lambda}$-enhancement factor can be only calculated. 
\begin{table*}
\centering
\begin{tabular}{|c|c|c|}            
\hline
Signal per Event ($\cdot\,10^{-4}$) & LVL1   & LVL2   \\
\hline
Data          &$ 0.17\pm 0.09 (stat)\substack{+0.02 \\ -0.06}$&$  2.36\pm 0.05 (stat)\substack{+0.23 \\ -0.31}$ \\
pp Simulation &$ 0.52\pm 0.09 (stat)\substack{+0.04 \\ -0.01}$&$  1.68\pm 0.36 (stat)\substack{+0.39 \\ -0.06}$   \\
\hline
\end{tabular}
\caption{Number of reconstructed $\mathrm{\Lambda}$ per event for the LVL1 and LVL2 event samples. The values are corrected for the no target events (see table~\ref{tab:triggerratio} and text for details).}
\label{table:lam:num1}
\end{table*}

In table~\ref{table:lam:num1} the number of reconstructed $\mathrm{\Lambda}$ candidates per event for the LVL1 and LVL2 samples are shown. The systematical errors were obtained by varying the reconstruction cuts within $\pm10\%$. 
One can see that the number of the reconstructed $\mathrm{\Lambda}$ in the experimental data and in the simulations is in agreement within the error. Possible differences might arise from the uncertainty of the total production cross-section in the simulations, hence one should focus more on the comparison of the ratios.
From these resulting values the enhancement is finally obtained by the division of the signal per event of LVL2 triggers for the value obtained from the LVL1 sample.
The LVL2 to LVL1 enhancement factor obtained in the simulation is $3.40\pm 0.59 (stat)\substack{+0.34 \\ -0.19}$ which is in quite good agreement with the enhancement of the pure geometrical acceptance (see section~\ref{sec:signal:back}). 

 The LVL2 to LVL1 enhancement obtained from the data is $14.1\pm 7.9 (stat)\substack{+4.3 \\ -0.6}$, where the huge statistical error is due to the scarce statistics of reconstructed $\mathrm{\Lambda}$ in the LVL1 sample. 
 This number is much larger than the value obtained for the simulation due to the fact that the simulations do not contain off-vertex reactions. If the simulation results are corrected for the fraction of off-vertex interactions in the LVL1 and LVL2 respectively, a $\mathrm{\Lambda}$ enhancement factor of $15.9\pm 2.8 (stat)\substack{+5.4 \\ -0.5}$ is obtained, which is in good agreement with the experimental data. 
 This way the performances of the LVL2 trigger have been proven and quantitatively verified.
\section{Summary}
\label{Summary}
We have presented the design, development and performance of  \silvioo. This device is composed of two layers of silicon detectors and was developed to function as an online trigger for $\mathrm{\Lambda}$ hyperon decay in $\mathrm{p+\pi^-}$ from p+p collisions at $3-4\,\mathrm{GeV}$.  \silvio was built to fit inside the FOPI spectrometer at GSI, closed to the target region and was utilized during an experimental campaign with the dedicated goal to look for the existence of kaonic bound states like $\mathrm{ppK^-}$ by analyzing the final state $\mathrm{p+p}$ $\mathrm{\rightarrow p+\Lambda+K^+}$. The silicon detectors were read out with two different systems. On the one hand side the analog Mesytec modules were employed to compute the multiplicity of the hits by charged particles on the two silicon layers and build the online trigger. On the other side, the p-side of the second layer of \silvio was read out with the front-end APV25 chip, to be able to handle the 480 strip signal in the offline analysis. The shown design resulted in a compact structure.
The efficiency and purity of the online trigger for the MIPs signal was extracted by measurements with beam and were discussed in this work. It resultes into an efficiency and purity of $97\,\%$ and at least $95\,\%$ (Multiplicity=1) respectively. The effect of the trigger device on empty-target events was studied,too. The LVL2 trigger, built by combining the FOPI LVL1 trigger with the \silvio multiplicity information, turned out to be also very efficient as a reaction trigger, since it could suppress empty-target events with a purity of $90\,\%$.\\
The momentum resolution of the FOPI spectrometer in the forward region is rather poor, the additional hit point provided by the \silvio B layer could improve the momentum determination of about $5\,\%$ for protons emitted in this phase space region. The \silvio hit information also improved the reconstruction of the secondary decay vertex for $\mathrm{\Lambda}$ hyperons, yielding to a reconstructed invariant mass resolution of $\mathrm{5.3\,MeV/c^2}$. This value extracted for the experimental data sample is comparable with the full-scale simulation.
The performance of the $\mathrm{\Lambda}$ trigger was evaluated first by means of the full-scale simulations of the reaction $\mathrm{p+p\rightarrow p+\Lambda+K^+}$ and of the background reactions excluding the $\mathrm{\Lambda}$ production. The background reduction achieved by employing the LVL1 and LVL2 trigger condition extracted with the simulations is determined to be $2.86 \pm 0.05 (stat) \pm 0.91 (ang)$ and  $11.93 \pm 0.12 (stat) \pm 3.7 (ang) $ respectively. These factors contain only the effect of the geometrical acceptance of the events, before the hyperon reconstruction. The enhancement of events containing a $\mathrm{\Lambda}$ candidate was studied in the experimental data comparing the hyperon reconstruction achieved for the LVL2 and LVL1 sample. Due to the limited statistics of the LVL1 sample, a large systematic error must be considered for the experimental ratio which was found to be $14.1\pm 7.9 (stat)\substack{+4.3 \\ -0.5}$. By including in the simulation the hyperon reconstruction and the  contribution by the empty-target events, a final value of $15.9\pm 2.8 (stat)\substack{+5.4 \\ -0.5}$ was obtained which is in very good agreement with the experimental findings. 
In this way the effectivity of \silvio was verified.
\section{Acknowledgements}
We would like to thank Robert Gernh\"auser and J\"urgen Friese for reading and commenting the manuscript.\\
This work was supported by the Helmholtz VH-NG-330, by the DFG cluster of excellence ʻOrigin and Structure of the
Universeʼ and the TUM Graduate School. Part of the work was supported by the Austrian Science Fund FWF Project P21457-N16.
\bibliographystyle{panda_tdr_lit}
\bibliography{literature}

\begin{flushleft}\begin{thebibliography}{10}\sloppy

\bibitem{DISTO}
M.~Maggiora~et. al.,
\newblock IEEE Trans. Nucl. Sci. {\bf 45}, 817 (1998).

\bibitem{COSYTOF}
M.~Dahmen et~al.,
\newblock Nucl. Instr. Meth. Phys. Ref. A {\bf 97} (1994).

\bibitem{FOPI1}
J.Ritman et~al.,
\newblock Nucl. Phys. Proc. Suppl. {\bf 44}, 708 (1995).

\bibitem{PROPOSAL}
Fopi-Collaboration,
\newblock Search for kaonic nuclear cluster $K^{-}pp$ in the $p + p \rightarrow
  K^+ + K^-pp \rightarrow K^+ + p \Lambda$ reaction with FOPI,
\newblock Experimental Proposal to GSI, 2007.

\bibitem{Yam1}
T.~Yamzaki and Y.~Akaishi,
\newblock Phys. Rev. C {\bf 65} (2002).

\bibitem{Yam2}
T.~Yamzaki and Y.~Akaishi,
\newblock Phys. Rev. C {\bf 76} (2007).

\bibitem{LOPEZ}
X.~Lopez~et al.,
\newblock Prog. Part. Nucl. Phys. {\bf 53}, 149 (2004).

\bibitem{FOPI2}
K.~Hildenbrand,
\newblock GSI Nachr. {\bf 91-02}, 6 (1992).

\bibitem{FOPI3}
A.~Gobbi~et al.,
\newblock NIM A {\bf 324} (1993).

\bibitem{RPC1}
A.~Sch\"uttauf,
\newblock NIM A {\bf 553}, 65 (2004).

\bibitem{RPC2}
A.~Sch\"uttauf et~al.,
\newblock Nuc. Phys. B {\bf 158}, 52 (2006).

\bibitem{beam}
P.~B\"uhler, O.~Hartmann, M.~Schafhauser, K.~Suzuki, and J.~Zmeskal,
\newblock Start counter and target system for the FOPI S349 experiment,
\newblock GSI Annual Report, 2009.

\bibitem{canb}
http://www.canberra.com.

\bibitem{mesytec}
http://www.mesytec.com.

\bibitem{APV3}
L.~Jones,
\newblock APV25-S1 - User Guide Version 2.2, 2001.

\bibitem{APV4}
N.~Raymond et~al.,
\newblock The CMS Tracker APV25 0,25 $\mu$m CMOS Readout Chip,
\newblock Paper presented at the 6th workshop on electronics for LHC
  experiments, Krakow, Poland, 2000.

\bibitem{gsiweb}
http://www.gsi.de.

\bibitem{bridgeboard}
M.~B\"ohmer,
\newblock {\em Messung der Spallationsreaktion $^{56}Fe+p$ in inverser
  Kinematik},
\newblock Dissertation, TU M\"unchen, 2006.

\bibitem{rmue}
R.~Muenzer,
\newblock Si$\Lambda$ViO - Ein Trigger fuer $\Lambda$-Hyperonen,
\newblock Diploma Thesis, 2008.

\bibitem{MLL}
Maier-Leibniz-Laboratoriums,
\newblock Homepage:,
\newblock http://www.bl.physik.uni-muenchen.de/.

\bibitem{plettner}
C.~Plettner,
\newblock {\em Strangenesspoduktion bei kleinen transversalen Impulsen und
  mittleren Rapiditäten in der Reaktion 96Ru +96Ru @ 1.69A.GeV},
\newblock Dissertation, Technische Universität Dresden, 1999, 1999.

\bibitem{wind}
H.~Wind,
\newblock Nucl. Instr. Meth. {\bf 115} (1974).

\bibitem{FOPILAM2}
any paper with fopi~lambda reconstruction.

\bibitem{Pluto}
I.~Froehlich et~al.,
\newblock ACAT2007 {\bf 76} (2007).

\bibitem{cosy3}
M.~e.~a. Abdel-Bary~et al.,
\newblock Eur.Phys.J. {\bf A46}, 27 (2010).

\bibitem{URQMD}
S.~Bass et~al.,
\newblock Prog. Part. Nucl. Phys. {\bf 41}, 225 (1998).

\end{thebibliography}\end{flushleft}
\end{document}